\documentclass[12pt,preprint]{aastex}

\slugcomment{Astronomical Journal}
\shorttitle{HCN/HCO+ in LIRGs}
\shortauthors{Imanishi et al.}

\begin{document}

\title{Millimeter Interferometric HCN(1--0) and HCO$^{+}$(1--0)
Observations of Luminous Infrared Galaxies} 

\author{Masatoshi Imanishi\altaffilmark{1}}
\affil{National Astronomical Observatory, 2-21-1, Osawa, Mitaka, Tokyo
181-8588, Japan} 
\email{masa.imanishi@nao.ac.jp} 

\author{Kouichiro Nakanishi\altaffilmark{1}}
\affil{Nobeyama Radio Observatory, Minamimaki, Minamisaku, Nagano,
384-1305, Japan} 

\author{Yoichi Tamura\altaffilmark{2}}
\affil{Department of Astronomy, The University of Tokyo, 7-3-1
Hongo, Bunkyo-ku, Tokyo 113-0033, Japan} 

\author{Nagisa Oi}
\affil{Department of Astronomy, School of Science, Graduate
University for Advanced Studies, Mitaka, Tokyo 181-8588}

\and

\author{Kotaro Kohno}
\affil{Institute of Astronomy, University of Tokyo, 2-21-1, Osawa, Mitaka, 
Tokyo, 181-0015, Japan}

\altaffiltext{1}{Department of Astronomy, School of Science, Graduate
University for Advanced Studies, Mitaka, Tokyo 181-8588}

\altaffiltext{2}{National Astronomical Observatory of Japan, 2-21-1
Osawa, Mitaka, Tokyo 181-8588, Japan} 

\begin{abstract}
We present the results on millimeter interferometric observations of
four luminous infrared galaxies (LIRGs), Arp 220, Mrk 231, IRAS
08572+3915, and VV 114, and one Wolf-Rayet galaxy, He 2--10, using 
the Nobeyama Millimeter Array (NMA). Both the HCN(1--0) and HCO$^{+}$(1--0)
molecular lines were observed simultaneously and their 
brightness-temperature ratios were derived.
High-quality infrared $L$-band (2.8--4.1 $\mu$m) spectra were also obtained
for the four LIRGs to better constrain their energy sources deeply
buried in dust and molecular gas.
When combined with other LIRGs we have previously observed with NMA,  
the final sample comprised nine LIRGs (12 LIRGs' nuclei) with available 
interferometric HCN(1--0) and HCO$^{+}$(1--0) data---sufficient to investigate
the overall trend in comparison with known AGNs and starburst galaxies.
We found that LIRGs with luminous buried AGN signatures at other
wavelengths tend to show high HCN(1--0)/HCO$^{+}$(1--0)
brightness-temperature ratios as seen in AGN-dominated galaxies,
while the Wolf-Rayet galaxy He 2--10 displays a small ratio. 
An enhanced HCN abundance in the interstellar gas surrounding a 
strongly X-ray-emitting AGN, as predicted by some chemical 
calculations, is a natural explanation of our results. 
\end{abstract}

\keywords{galaxies: active --- galaxies: nuclei ---  galaxies: ISM ---
radio lines: galaxies --- galaxies: individual (\objectname{Arp 220, 
Mrk 231, IRAS 08572+3915, VV 114 and He 2--10})} 

\section{Introduction}

It has recently been found that supermassive blackholes (SMBH) with
masses of M$_{\rm SMBH}$ $>$ 10$^{6}$M$_{\odot}$ are ubiquitously
present at the centers of spheroidal components (bulges and elliptical
galaxies) \citep{mag98}. 
When active mass accretion onto a SMBH occurs, it is observed as an
active galactic nucleus (AGN). 
The current AGN picture postulates that dust is present in
torus geometry around the central SMBH \citep{ant93}. 
The direction perpendicular to the torus is mostly transparent to the
bulk of the AGN's ionizing UV -- soft X-ray photons. 
Gas clouds at 10--1000 pc distances from the AGN along the torus axis,
photo-ionized by the central AGN's radiation, produce the so-called
narrow-line regions (NLRs; Robson 1996). 
Since the spectral shapes of the ionizing photons differ between
AGNs and stars, emission-line ratios from NLRs differ from gas
clouds photo-ionized in star-forming galaxies.
Thus, AGNs surrounded by torus-shaped dust with well developed NLRs are 
distinguishable from star-forming galaxies relatively easily
\citep{vei87}. 

Because dust at the central $<$10 pc of galaxies has an angular momentum,
the spatial distribution of the dust is likely to be axisymmetrical; 
dust column density is higher in one direction (the torus
direction) and lower in another direction (the torus axis). 
When the nuclear concentration of dust becomes large, 
even the torus axis direction can be opaque to the bulk of the AGN's 
UV -- soft X-ray radiation, and the radiation can be blocked at the
inner part ($<$10 pc), producing virtually no NLRs.  
Such {\it buried} AGNs are no longer detectable as long as emission
lines originating in the NLRs are sought. 
Because a simple interpretation of the cosmic X-ray background spectrum 
suggests that most AGNs in the universe are buried \citep{fab99}, 
it is important to establish a technique for finding buried AGNs. 
In a buried AGN, almost all of the energetic UV -- soft X-ray radiation
is absorbed by the surrounding dust and is re-emitted as infrared dust
emission.
For this reason, {\it luminous} buried AGNs are expected to be strong
infrared emitters. 
Luminous infrared galaxies (LIRGs), which radiate strong infrared 
emission with L$_{\rm IR}$ $>$ 10$^{11}$L$_{\odot}$ and contain 
highly concentrated nuclear dust \citep{sam96}, are probable objects in
which luminous buried AGNs reside. 

Since star-forming activity can also produce infrared emission, it is
necessary to distinguish the origin of the infrared luminosities of
LIRGs if we are to study the luminous buried AGN population.
An effective method is to search for the presence of strong X-ray
emission because an AGN intrinsically produces much stronger X-ray
emission compared to star-forming activity \citep{rob96}. 
Unfortunately, buried AGNs generally contain a large amount of nuclear 
dust \citep{idm06}, and most of them may be Compton-thick 
(N$_{\rm H}$ $>$ 10$^{24}$ cm$^{-2}$) sources \citep{mai03}. 
X-ray observations at E $>$ 10 keV are needed to directly detect Compton-thick
X-ray emission from luminous buried AGNs with a small
scattered/reflected component \citep{fab02}; however, these are
applicable only to a few very nearby bright LIRGs due to a current
sensitivity limit. 

Although a systematic buried AGN search through direct X-ray
observations is difficult, we can investigate the presence of 
a luminous buried AGN through the chemical effects of the intrinsically
strong X-ray emission on the surrounding interstellar
medium. 
Around a strongly X-ray-emitting buried AGN surrounded by dense
gas and dust, the so-called X-ray dissociation regions (XDR; Maloney et
al. 1996) should develop, in contrast to photo-dissociation regions
(PDRs) usually seen in strongly UV-emitting star-forming regions. 
Detection of XDR signatures can thus be a useful tool for locating a luminous
buried AGN. 
Among the several proposed probes of XDRs \citep{use04,aal07}, we focus 
on the {\it observationally derived} method based on HCN(1--0) and 
HCO$^{+}$(1--0) line ratios \citep{koh01,koh05,ink06,gra06}, because
theoretical prediction of line fluxes from PDRs and XDRs requires 
many free parameters \citep{yam07} and could still be 
uncertain, given that the most important parameter---detailed
properties of the clumpy structure of molecular gas \citep{sol87}---is 
totally unknown from observations.     

Here, we present the results of our millimeter interferometric 
HCN(1--0) and HCO$^{+}$(1--0) observations using the Nobeyama
Millimeter Array (NMA) of nearby LIRGs whose energy sources have 
previously been investigated at other wavelengths. 
Compared to previous HCN(1--0) and HCO$^{+}$(1--0) observations of LIRGs
using single-dish radio telescopes \citep{gao04,gra06}, our interferometric
maps have the important advantage that spatially resolved 
HCN(1--0) and HCO$^{+}$(1--0) data are obtainable in LIRGs, which often show 
disturbed, multiple nuclei morphologies \citep{sam96}.
New ground-based infrared $L$-band (2.8--4.1 $\mu$m) spectra of higher
quality than older data were also taken for individual main nuclei of these
millimeter-observed LIRGs to better constrain the nature of their
obscured energy sources. 
We adopted H$_{0}$ $=$ 75 km s$^{-1}$ Mpc$^{-1}$,
$\Omega_{\rm M}$ = 0.3, and $\Omega_{\rm \Lambda}$ = 0.7 throughout this
paper.  

\section{Targets}

The target LIRGs are selected on the basis of  
their proximity ($z <$ 0.06) to cover the redshifted
HCN(1--0) and HCO$^{+}$(1--0) emission lines with the NMA receiving
systems, and  their expected high fluxes of HCN(1--0) and HCO$^{+}$(1--0).
The four LIRGs, Arp 220, IRAS 08572+3915, Mrk 231, and VV 114, were observed.
Table 1 summarizes the infrared emission properties of these LIRGs. 
An angular scale of 1$''$ corresponds to a physical size of 0.3--1 kpc
at redshift = 0.018--0.058. 

Arp 220 (z = 0.018) is one of the best-studied nearby ultraluminous 
infrared galaxies (ULIRGs; L$_{\rm IR}$ $>$ 10$^{12}$L$_{\odot}$; Sanders
et al. 1988a). 
It consists of two nuclei, Arp 220 E and W, with a separation of
$\sim$1$''$ \citep{sco00,soi00}, and the optical spectrum of the combined
emission is classified as a LINER (i.e., non-Seyfert; no obvious AGN
signatures in the optical spectrum) \citep{vei99}.
The presence of starburst activity (= active star-formation) is
evident in various observational data \citep{gen98,imd00}, but the
detected starburst is energetically insufficient to fully account for
the observed infrared luminosity of Arp 220 quantitatively 
\citep{idm06,arm07,ima07}, requiring the energy source deeply buried in
Arp 220's nuclear core \citep{spo04,gon04}.  
The presence of a luminous buried AGN at the core has long been unclear 
even through detailed observations in the X-ray \citep{iwa05}, infrared 
\citep{gen98,arm07}, and radio \citep{shi01,par07} frequencies. However,
the recent discovery of a compact energy source with a very high
emission surface brightness has provided the first strong signatures of
a luminous buried AGN in Arp 220W \citep{dow07}.

Mrk 231 (z = 0.042) is a well studied, single nucleus ULIRG and is the
most luminous object (L$_{\rm IR}$ $\sim$ 10$^{12.5}$L$_{\odot}$) in the
local universe. 
It is an optically known AGN classified as a Seyfert 1 galaxy
\citep{vei99}, but it displays absorption features in the infrared
spectra \citep{roc83,arm07} and X-rays \citep{bra04} possibly
originating in the broad absorption line (BAL) clouds \citep{bok77}. 
The presence of a very luminous and probably energetically dominant AGN is
revealed by various observations (e.g., Soifer et al. 2000; 
Imanishi \& Dudley 2000; Braito et al. 2004; Imanishi et al. 2006a). 

IRAS 08572+3915 (z = 0.058) is a ULIRG (L$_{\rm IR}$ $\sim$
10$^{12.1}$L$_{\odot}$) consisting of two nuclei (NW and SE) with a
separation of $\sim$5$''$ \citep{sco00,kim02}. 
The NW nucleus (IRAS 08572+3915NW) is believed to be energetically
dominant \citep{soi00} and is therefore our primary interest. 
It is classified optically as a LINER \citep{vei99}. It is one of the
strong {\it buried} AGN candidates because (1) the energy source is
suggested to be very compact and is more centrally concentrated than the
nuclear dust, as expected for a buried AGN
\citep{dud97,soi00,idm06,ima07}, and (2) the polycyclic aromatic
hydrocarbons (PAH) emission (the starburst indicator) is extremely weak
\citep{imd00,idm06,spo06,arm07,ima07}.  
It is thus a particularly interesting target for investigating whether the
HCN(1--0) to HCO$^{+}$(1--0) brightness-temperature ratio is 
similar to those found in AGN-dominated nuclei or starburst-dominated
galaxies \citep{koh05}. 

VV 114 (z = 0.020) is a double-nuclei LIRG (L$_{\rm IR}$ $\sim$
10$^{11.6}$L$_{\odot}$) with a separation of $\sim$15$''$ 
\citep{kno94}.
Both nuclei are classified optically as HII regions \citep{vei95}. 
In the short optical wavelength, the western nucleus (VV 114W) is
brighter than the eastern nucleus (VV 114E), but VV 114E becomes
more important with increasing wavelength and is prominent in the
infrared band \citep{kno94,lef02}, probably dominating the infrared
luminosity of the VV 114 merging system. 
The VV 114E nucleus has a secondary peak $\sim$1$\farcs$5 southwest of
the VV 114E peak (VV 114E$_{\rm SW}$; Soifer et al. 2001).
The presence of a luminous buried AGN is suggested in VV 114E from
the strong featureless mid-infrared 15 $\mu$m continuum emission
\citep{lef02}. 

In Kohno (2005), HCN(1--0) to HCO$^{+}$(1--0) brightness-temperature
ratios are studied in AGNs and typical starburst galaxies such as M82. 
In the nuclei of LIRGs, where molecular gas is very highly concentrated
\citep{sam96}, putative star formation could be extreme in that
young, massive, hot stars are more predominant than a normal starburst. 
To determine if such an extreme starburst shows a different
HCN(1--0)/HCO$^{+}$(1--0) brightness-temperature ratio from a normal
starburst, a nearby well studied Wolf-Rayet galaxy, He 2--10 (z =
0.003), dominated by massive, hot Wolf-Rayet stars \citep{all76} was
also studied (Table 1).   

\section{Observations and Data Reduction}

\subsection{Millimeter interferometry}

Interferometric observations of HCN(1--0) ($\lambda_{\rm rest}$ =
3.3848 mm and $\nu_{\rm rest}$ = 88.632 GHz) and HCO$^{+}$(1--0)
($\lambda_{\rm rest}$ = 3.3637 mm or $\nu_{\rm rest}$ = 89.188 GHz)
lines were undertaken with the Nobeyama Millimeter Array (NMA) and the 
RAINBOW interferometer at the Nobeyama Radio Observatory (NRO) between
2005 January and 2007 April. 
Table 2 summarizes the detailed observation log. 
The NMA consists of six 10-m antennas and observations were undertaken
using the AB (the longest baseline was 351 m), C (163 m), and D (82 m)
configurations. The RAINBOW interferometer is a seven-element combined 
array that includes the NRO 45-m telescope in addition to the six 10-m
antennas (NMA). 
The RAINBOW observations were scheduled when the NMA array was set at
the AB configuration and the longest baseline was 410 m. 
In the 3-mm wavelength range, the sensitivity of the RAINBOW interferometer 
is about twice better than that of the NMA array only because the total
aperture size increases by a factor of four with the inclusion of the
NRO 45-m telescope. 

The backend was the Ultra-Wide-Band Correlator (UWBC) \citep{oku00}
configured to cover 1024 MHz with 128 channels at 8-MHz
resolution. The central frequency for each source (Table 2) was set to 
cover both the redshifted HCN(1--0) and HCO$^{+}$(1--0) lines
simultaneously.  
A bandwidth of 1024 MHz corresponds to $\sim$3500 km s$^{-1}$ at $\nu$
$\sim$ 84--89 GHz. The field of view at this frequency is $\sim$26$''$
(full-width at half-maximum; FWHM) and $\sim$77$''$ (FWHM) for the RAINBOW
and the NMA array, respectively.   
The Hanning window function was applied to reduce side lobes in
the spectra; thus, the effective resolution was widened to 16 MHz or 55 km
s$^{-1}$ at $\nu$ $\sim$ 84--89 GHz. 
Since the declinations of VV 114 and He 2--10 are low ($<$ $-$15$^{\circ}$), 
an observable time period (elevation $>$ 20$^{\circ}$) per each day from
the NMA site is short, requiring many observing days.

The UVPROC-II package developed at NRO \citep{tsu97} and the AIPS
package of the National Radio Astronomy Observatory were used
for standard data reduction. Corrections for the antenna baselines,
band-pass properties, and the time variation in the visibility amplitude
and phase were applied to all of the data (Table 2). A fraction of the
data with large phase scatter due to poor millimeter seeing was removed
from our analysis. After clipping a small fraction of data of unusually
high amplitude, the data were Fourier-transformed using a natural 
{\it UV} weighting. The flux calibration was made using 
observations of Uranus or appropriate quasars (Table 2) whose flux
levels relative to Uranus or Neptune had been measured at least every
month in the NMA observing seasons. 
A conventional CLEAN method was applied to deconvolve the synthesized
beam pattern. 
The primary beam pattern of the NMA antenna was corrected for LIRGs with
very extended spatial structures (VV 114 and He 2--10). 
Table 3 summarizes the total net on-source integration times and
synthesized beam patterns. 
Absolute positional uncertainties of the NMA/RAINBOW maps were estimated to 
be much less than 1$''$.

\subsection{Infrared 2.8--4.1 $\mu$m ($L$-band) spectroscopy}

We conducted ground-based infrared 2.8--4.1 $\mu$m ($L$-band) slit 
spectroscopy on the main nuclei of the four LIRGs (Arp 220, Mrk 231, 
IRAS 08572+3915, and VV 114) to better understand the nature of their 
obscured energy sources based on the equivalent width of the 
3.3 $\mu$m PAH emission and the optical depths of absorption features at
$\lambda_{\rm rest}$ $\sim$ 3.05 $\mu$m (in the rest frame) by
ice-covered dust grains and at $\lambda_{\rm rest}$ $\sim$ 3.4 $\mu$m by
bare carbonaceous dust grains \citep{imd00,imm03}. 
A normal starburst galaxy should always show large
equivalent-width PAH emission, while a pure AGN produces a PAH-free
continuum \citep{imd00,idm06}.
For absorption features, optical depths have upper limits in a
normal starburst in which stellar energy sources and dust are spatially
well mixed, while they can be arbitrarily large in a buried AGN with a more 
centrally concentrated energy source geometry than dust 
\citep{imm03,idm06}. 
For these reasons, these PAH emission and dust-absorption features 
can be used to distinguish the dust-obscured energy sources of LIRGs. 

Ground-based infrared spectra taken with large ($>$4-m)
telescopes are superior in spatial resolution to those obtained
by space satellites with small diameters ($<$1 m), thus enabling us
to obtain spatially resolved spectra of double nuclei with a small
separation ($<$2 arcsec).
For example, Arp 220 and VV 114E have double nuclei with separations of 
1 to 1.5 arcsec, which can be resolved only with the ground-based
spectra.
For Mrk 231 and IRAS 08572+3915, although infrared $L$-band spectra taken
with $<$4m telescopes have been published 
\citep{ima98,imd00,idm06}, a higher-quality $L$-band spectrum
obtained with a larger, 8-m-class telescope may be able to provide new
constraints on their energy sources. 
We thus obtained infrared $L$-band spectra of Arp 220, Mrk 231, 
IRAS 08572+3915, and VV 114E using the Subaru 8.2-m telescope
\citep{iye04} atop Mauna Kea, Hawaii.

We used the IRCS infrared spectrograph \citep{kob00} at the 
Nasmyth focus of the Subaru telescope to obtain the new
infrared $L$-band spectra of these LIRGs. 
Table 4 tabulates the observation log. 
The sky was clear during the observations. The seeing in the $K$-band
measured in images taken before the $L$-band spectroscopy was
0$\farcs$4--0$\farcs$8 in FWHM. 
A 0$\farcs$6- or 0$\farcs$9-wide slit and the $L$-grism were used with a
52-mas-pixel scale. 
The achievable spectral resolution is R $\sim$ 140--200 at $\lambda
\sim$ 3.5 $\mu$m. 
When the seeing size was sufficiently small, we basically used the
0$\farcs$6 slit to search for AGN signatures at a nuclear core 
based on the PAH equivalent width and the strengths of dust absorption
features \citep{imd00,imm03,idm06} with a reduced contamination from
extended starburst components.
However, for IRAS 08572+3915, the 0$\farcs$9 slit was employed because
the seeing in the $K$-band worsened during the observation. 
Thus, PAH emission from an extended ($>$0$\farcs$6 or $>$0$\farcs$9)
starburst component is not covered; thus, PAH flux or luminosity is not
considered in our discussion. 
For Arp 220, Mrk 231, and IRAS 08572+3915, nuclear PAH fluxes (or 
luminosities) within the central 1--2 arcsec can better be investigated from
previously obtained spectra using a wider slit \citep{imd00}.
The precipitable water was low, 1--2 mm for the 2006 July observing
run, and $<$1 mm for the 2007 January run. 
We employed a standard telescope nodding technique (ABBA pattern) with a
throw of 5 to 7 arcsec along the slit to subtract background emission.
We used the optical guider of the Subaru telescope to monitor the
telescope tracking. 
Exposure time was 1.0--1.5 sec, and 30--60 coadds were made at each nod
position. 
            
For both observing runs, F- or G-type main sequence stars
(Table 4) were observed as standard stars, with a mean airmass
difference of $<$0.1 to the individual LIRG nuclei to correct for
the transmission of Earth's atmosphere and to provide flux
calibration. The $L$-band magnitudes of the standard stars were
estimated from their $V$-band ($\lambda =$ 0.6 $\mu$m) magnitudes,
adopting the $V-L$ colors appropriate to the stellar types of
individual standard stars \citep{tok00}.

Standard data analysis procedures were employed using IRAF   
\footnote{IRAF is distributed by the National Optical Astronomy
Observatories, which are operated by the Association of Universities
for Research in Astronomy, Inc. (AURA), under cooperative agreement
with the National Science Foundation.}.
Initially, frames taken with an A (or B) beam were subtracted from
frames subsequently taken with a B (or A) beam, and the resulting
subtracted frames were added and divided by a spectroscopic flat
image. Bad pixels and pixels hit by cosmic rays were then replaced
with the interpolated values of the surrounding pixels. Finally, the
spectra of the LIRGs' nuclei and standard stars were extracted by
integrating signals over 0$\farcs$8--1$\farcs$8, depending on the
actual signal profiles. 
To create a combined spectrum of Arp 220E+W, 
we summed signals over 2$\farcs$5--2$\farcs$9 along the slit. 
Wavelength calibration was performed based on the wavelength-dependent
transmission of Earth's atmosphere. 
The spectra of the LIRGs' nuclei were divided by the observed spectra of
standard stars and multiplied by the spectra of blackbodies with
temperatures appropriate to individual standard stars (Table 4).

Flux calibration was performed based on signals of the LIRGs and standard stars
detected inside our slit spectra. 
Seeing sizes in the $K$-band (and $L$-band) were always smaller
than the employed slit widths, and good telescope tracking performance
of Subaru was established. 
We thus expected possible slit loss for the compact ($<$0$\farcs$6)
emission to be minimal, say $<$30\%. 
In fact, we divided the whole data set into subgroups and compared
their flux levels, which agreed within 20\% for all LIRGs.
This level of possible flux uncertainty ($<$30\%) for the compact
emission will not seriously affect our main conclusions.
To reduce scatter of the infrared spectra, appropriate binning of spectral
elements was performed, particularly at $\lambda_{\rm obs} < 3.3$ $\mu$m
in the observed frame, where the scatter of data points is larger due
to poorer Earth atmospheric transmission than at $\lambda_{\rm obs} >
3.4$ $\mu$m. 

\section{Results}

\subsection{Millimeter interferometric data}

For Arp 220 and Mrk 231, millimeter spectra at HCN(1--0) or
HCO$^{+}$(1--0) emission peak positions show that the flux levels
between these lines are substantially above the zero level, indicating
that continuum emission is clearly present. 
We combined data points that are unaffected by these lines and made
interferometric maps of the continuum emission. 
Figure 1 presents the contours of the continuum emission. 
The continuum emission is clearly ($>$9$\sigma$) detected in Arp 220 and
Mrk 231.
We barely see the continuum emission signature for VV 114, IRAS
08572+3915, and He 2--10, but its detection significance is $<$3$\sigma$.
Table 5 presents the estimated continuum levels.

Figure 2 shows the integrated intensity maps of the HCN(1--0) and
HCO$^{+}$(1--0) emission of the five observed sources  
(Arp 220, Mrk 231, IRAS 08572+3915, VV 114, and He 2--10). 
The continuum emission is subtracted for Arp 220, Mrk 231, and VV 114, 
but not for IRAS 08572+3915 and He 2--10 because of possible large 
ambiguities. 

Since molecular gas is highly concentrated at the nuclei of LIRGs, the
importance of high-density molecular gas (n$_{\rm H}$ $>$ 10$^{4}$
cm$^{-3}$) is expected to increase there \citep{gao04}.  
Such molecular gas is better probed with HCN(1--0) and HCO$^{+}$(1--0)
rather than the widely used CO(1--0) because the dipole moments of 
HCN and HCO$^{+}$ ($\mu$ $>$ 3 debye) are much larger than CO 
($\mu$ $\sim$ 0.1 debye; Botschwina et al. 1993; Millar et al. 1997).
Hence, the maps in Figure 2 reflect the spatial distribution of
this important dense molecular gas.
A spatially unresolved emission peak is clearly seen for Mrk 231 at the
nuclear position. 
For IRAS 08572+3915, the spatially unresolved HCN(1--0) emission peaks at
the energetically dominant NW nucleus, as expected (see $\S$ 2,
paragraph 4).  
For He 2--10, HCO$^{+}$(1--0) emission displaying no significant spatial
extent is found close to the CO(1--0) peak \citep{kob95}.
A spatially extended emission component is detected for Arp 220 and VV 114.
Figures 3 and 4 display the channel maps around the HCN(1--0) and
HCO$^{+}$(1--0) lines of Arp 220 and VV 114, respectively.
 
For Arp 220, the HCN(1--0) and HCO$^{+}$(1--0) emission
peaks reside in between the double nuclei in the integrated intensity
maps of Figure 2. Nevertheless, the channel map in Figure 3 clearly
shows that Arp 220E becomes more prominent with increasing velocity
(upper-left, red component) and Arp 220W does with decreasing velocity
(lower-right, blue component).  
Our HCN(1--0) and HCO$^{+}$(1--0) channel maps thus confirm that 
the Arp 220E nucleus is farther redshifted than Arp 220W, as was 
previously noted \citep{lar95,dow98,tan98,sak99}.

In the case of VV 114, the HCN(1--0) the emission peak is close to VV 114E,
while that of HCO$^{+}$(1--0) is strong between the VV 114E and W nuclei 
(Figure 2), similar to CO(1--0), (2--1), and (3--2) emission   
\citep{yun94,ion04}.  
The spatial distributions of HCN(1--0) and HCO$^{+}$(1--0) are quite
different, possibly because (a) HCN(1-0) and HCO$^{+}$(1--0) probe
molecular gas with different density \citep{gre06}, due to their
slightly different critical densities 
\footnote{
The critical density (n$_{\rm c}$) is, by definition, proportional to 
Einstein's A coefficient (A) divided by the collision rates \citep{sch05}. 
The A-coefficient A is $\propto$ $\mu^{2}$ $\times$ $\nu^{3}$, where 
$\mu$ is the dipole moment and $\nu$ is the frequency of the molecular line. 
The dipole moments ($\mu$) of HCN ($\mu$ $\sim$ 3 debye; Millar et al. 1997) 
and HCO$^{+}$ ($\mu$ $\sim$ 4 debye; Botschwina et al. 1993) are similar
to each other. 
The frequencies of HCN(1--0) (88.632 GHz) and HCO$^{+}$(1--0) (89.188 GHz)
are almost identical.
Thus, the A-coefficient is similar between HCN(1--0) and HCO$^{+}$(1--0). 
However, using the collision rates estimated by \citet{sch05}, 
the n$_{\rm c}$ value of HCN(1--0) is a factor of $\sim$5 larger than 
that of HCO$^{+}$(1--0).
}, 
or (b) HCO$^{+}$ (1--0) may selectively trace  
shock regions, due to enhanced HCO$^{+}$ abundance by shocks 
\citep{dic80}.  
Further details will be discussed in a future paper 
(Tamura et al., in preparation). No significant HCN(1--0) or
HCO$^{+}$(1--0) emission is recognizable in VV 114W.

Figure 5 shows spectra around the HCN(1--0) and HCO$^{+}$(1--0)
lines for Arp 220, Mrk 231, IRAS 08572+3915, and VV 114.  
For Mrk 231 and IRAS 08572+3915, spectra are extracted at the 
nuclear peak position. 
For Arp 220, the spectra are extracted at both the Arp 220E and W
positions, and double-peaked HCN(1--0) and HCO$^{+}$(1--0)
emission is seen at both positions. 
Although the double-peaked profile suggests that emission from Arp
220E and W are not clearly spatially resolved in a map, 
the redshifted (lower frequency) component is stronger 
at the Arp 220E position than at Arp 220W, as expected from the channel
map (Figure 3).
We can thus separate the emission from Arp 220E and W spectroscopically,
with the red and blue components coming from Arp 220E and W,
respectively. 
For VV 114, both the HCN(1--0) and HCO$^{+}$(1--0) emission have 
spatial structures, and their peak positions are not the same. 
We thus extract spectra at four different positions: 
(E-1) the HCN(1--0) peak at the eastern side of VV 114E in Figure 2; 
(E-2) the HCN(1--0) peak at the western side of VV 114E in Figure 2;
(3) the eastern HCO$^{+}$(1--0) peak between VV 114E and W in Figure 2; and 
(4) the western HCO$^{+}$(1--0) peak between VV 114E and W in Figure 2.

Figure 6 presents Gaussian fits to the detected HCN(1--0) and 
HCO$^{+}$(1--0) lines.
The central velocity and line width of the Gaussian fit are determined
independently for the HCN(1--0) and HCO$^{+}$(1--0) lines. 
For IRAS 08572+3915, a double Gaussian fit was 
attempted because the lines seem to be double-peaked. 
Table 6 summarizes the Gaussian fitting results.
For Mrk 231 and IRAS 08572+3915 (the LIRGs dominated by the single
primary nucleus), the HCN(1--0) peak velocity is similar to that of
CO(1--0) in previous reports \citep{dow98,bry96,eva02}. 
For Arp 220E and W nuclei, the peak velocities of HCN(1--0) and
HCO$^{+}$(1--0) agree with that of CO(2--1) within 50 km s$^{-1}$
\citep{dow98}. 

For Arp 220, Mrk 231, IRAS 08572+3915, and VV 114, the integrated fluxes
of HCN(1--0) and HCO$^{+}$(1--0) at each peak position estimated from
the Gaussian fits in the spectra are summarized in Table 7.  
For He 2--10, the fluxes are derived from the peak contours of the
integrated intensity maps (Figure 2) because the spectrum is noisy. 
The HCN(1--0)/HCO$^{+}$(1--0) brightness-temperature ratios ($\propto$
$\lambda^{2}$ $\times$ flux density) are also shown in Table 7.
In Mrk 231 and IRAS 08572+3915, both the HCN(1--0) and HCO$^{+}$(1--0) 
emission show a spatially unresolved single peak at the main nuclear
position. 
The integrated HCN(1--0) and HCO$^{+}$(1--0) fluxes are also estimated 
from the peak contours of the integrated intensity maps
in Figure 2. The estimated values are consistent with those based
on the Gaussian fits for both HCN(1--0) and HCO$^{+}$(1--0) emission
within 15\% in Mrk 231 and 30\% in IRAS 08572+3915 if the continuum in
Figure 6 is assumed. 

Arp 220, VV 114, and He 2--10 display spatially extended 
HCN(1--0) and HCO$^{+}$(1--0) emission. 
Their total fluxes are shown in Table 8. 
For Arp 220 and Mrk 231, HCN(1--0) and HCO$^{+}$(1--0) fluxes 
have previously been measured using single-dish radio telescopes. 
Table 9 compares their fluxes with our measurements, which are similar,
suggesting that our interferometric data recover the bulk of HCN(1--0)
and HCO$^{+}$(1--0) emission.  

\subsection{Infrared spectra}

\subsubsection{PAH emission}

Figure 7 presents the flux-calibrated infrared 2.8--4.1 $\mu$m ($L$-band)
spectra of Arp 220, Mrk 231, IRAS 08572+3915NW, and VV 114E.
For Arp 220 and VV 114E, our ground-based spectra provide 
spatially resolved spectra of double nuclei with small (1--2 arcsec)
separation, which is not possible for space-based infrared spectra (see 
$\S$3.2).
Thus, our $L$-band spectra make it possible to discuss the energy sources
of individual nuclei separately for Arp 220E and W \citep{sco00,soi00},
and VV 114E and E$_{\rm SW}$ \citep{soi01}. 

The 3.3 $\mu$m PAH emission is clearly seen in the spectra of the
observed LIRGs, except for IRAS 08572+3915NW, indicating that a detectable
amount of starburst activity is surely present.
Table 10 summarizes the strength of the 3.3 $\mu$m PAH emission
feature.
The observed 3.3 $\mu$m PAH luminosities (L$_{\rm 3.3PAH}$) roughly trace
the absolute magnitudes of modestly obscured (A$_{\rm V}$ $<$ 15 mag)
starburst activities covered inside our slit spectra 
(the L$_{\rm IR}$ of such a starburst is $\sim$10$^{3}$ $\times$ 
L$_{\rm 3.3PAH}$; Mouri et al. 1990; Imanishi 2002).

For Arp 220E+W, Mrk 231, and IRAS 08572+3915NW, the equivalent widths of
the 3.3 $\mu$m PAH emission (EW$_{\rm 3.3PAH}$) are similar to previous
estimates based on old, lower-quality spectra \citep{imd00,idm06}.  
The EW$_{\rm 3.3PAH}$ values of Mrk 231, IRAS 08572+3915NW, and VV 114E
are more than a factor of $\sim$5 smaller than the typical value found
in starburst-dominated galaxies (EW$_{\rm 3.3PAH}$ $\sim$ 100 nm;
Imanishi \& Dudley 2000), indicating that PAH-free continua from the
putative luminous AGNs contribute importantly to their observed $L$-band
fluxes \citep{imd00,idm06,ima06}. 
For Arp 220E, W, and VV 114E$_{\rm SW}$, the EW$_{\rm 3.3PAH}$ values
are as large as seen in normal starburst galaxies, so that no 
explicit signs of PAH-free continua from buried AGNs are detected 
in this wavelength range.

For Mrk 231 and IRAS 08572+3915, the ULIRGs dominated by compact nuclear
emission, the $L$-band continuum flux levels in our slit spectra are
similar, within 0.1 mag, to the photometric measurements using
a 2$\farcs$5 aperture \citep{zho93}.
For Arp 220E+W, the $L$-band continuum flux level is a factor of $\sim$2
smaller than previously obtained spectra using a 1$\farcs$2 aperture
\citep{imd00} and photometric data using a 2$\farcs$5 aperture
\citep{zho93}. 
The small flux is possibly due to the small slit width (0$\farcs$6)
in the new spectra and/or possible $L$-band flux ambiguity in the
adopted standard star, HR 5728 (Table 4). 

\subsubsection{Absorption features}

IRAS 08572+3915NW (Figure 7) shows a strong absorption feature at
$\lambda_{\rm obs}$ = 3.6 $\mu$m in the observed frame or 
$\lambda_{\rm rest}$ = 3.4 $\mu$m in the rest frame due to bare
carbonaceous dust grains. 
Its optical depth is estimated to be $\tau_{3.4}$ $\sim$ 0.8, which is
similar to previous estimates \citep{imd00,idm06}. 
The optical depth is larger than the upper limit produced with the
natural geometry of a normal starburst (stellar energy sources and dust
spatially well mixed; $\tau_{3.4}$ $<$ 0.2), but requires buried
AGN-type geometry in which the energy source is more centrally concentrated
than the dust \citep{imm03,idm06}, supporting the buried-AGN classification
as previously indicated from the low EW$_{\rm 3.3PAH}$ value ($<$3 nm;
Table 10). 

In the high-quality IRAS 08572+3915NW spectrum in Figure 7, the broad
absorption feature centered at $\lambda_{\rm obs}$ = 3.2 $\mu$m or 
$\lambda_{\rm rest}$ = 3.0--3.05 $\mu$m is seen for the first time. 
In a buried AGN and starburst composite galaxy, the buried AGN emission 
is more highly obscured, because the central compact buried AGN 
should be geometrically surrounded by the starburst. 
Hence, the contribution from buried AGN emission to an observed flux 
becomes higher with increasing wavelength, from the $K$-band (2--2.5
$\mu$m) to $L$-band (2.8--4.1 $\mu$m), due to smaller dust extinction.
Additionally, blackbody radiation from stars (T $>$ 4000K) generally
shows a steeply decreasing flux from the $K$- to $L$-band.
For these reasons, it could happen that an observed $K$-band continuum
emission largely comes from foreground stellar emission, in addition to 
buried AGN emission, while an $L$-band continuum, particularly at a
longer wavelength part, is dominated by buried AGN emission.  
If this is the case for IRAS 08572+3915NW and if the stellar emission's
tail extends from the K-band to the shorter part of the 
$L$-band ($\lambda_{\rm obs}$ $<$ 3.0 $\mu$m), then an absorption-like
feature at $\lambda_{\rm rest}$ $\sim$ 3.05 $\mu$m could be reproduced. 
However, the $K$-band spectrum of IRAS 08572+3915
comes mostly from buried-AGN-heated hot dust emission, because of 
the extremely red $K$-band continuum and no stellar CO absorption
features at $\lambda_{\rm rest}$ $\sim$ 2.3 $\mu$m \citep{gol95,idm06}.
Thus, this scenario seems implausible. 
We ascribe the absorption-like feature to the 3.05 $\mu$m absorption by
ice-covered dust grains and estimate its optical depth to be
$\tau_{3.1}$ $\sim$ 0.3. 

The spectrum of VV 114E also displays the 3.05 $\mu$m ice-absorption
feature with an optical depth of $\tau_{3.1}$ $\sim$ 0.5. 
Its optical depth is again larger than the threshold explained by 
the mixed dust/source geometry of a normal starburst 
($\tau_{3.1}$ $<$ 0.3; Imanishi \& Maloney 2003), suggesting the
presence of a luminous buried AGN, as inferred from the low EW$_{\rm
3.3PAH}$ value ($\sim$20 nm; Table 10).  

For the infrared $L$-band spectra of the remaining LIRG nuclei (Arp
220E, W, Mrk 231, and VV 114E$_{\rm SW}$), no clear absorption features
are recognizable. 
The absence of absorption features in Mrk 231 (the low EW$_{\rm 3.3PAH}$
object; Table 10) implies that the AGN is only weakly obscured, while 
those for Arp 220E, W, and VV 114E$_{\rm SW}$ are explained by
the predominant contribution from starburst emission to the observed
$L$-band fluxes.  
For Arp 220W, the nucleus with luminous buried AGN signatures
\citep{dow07}, we see no explicit AGN sign in the infrared $L$-band
spectrum. 
The emission from the putative buried AGN may be so highly flux-attenuated
that its contribution to an observed $L$-band flux must be
insignificant, compared to foreground weakly-obscured starburst emission.

\section{Discussion}

\subsection{Comparison of HCN(1--0)/HCO$^{+}$(1--0)
brightness-temperature ratios with other galaxies}

Figure 8 shows the HCN(1--0)/HCO$^{+}$(1--0) and HCN(1--0)/CO(1--0)
brightness-temperature ratios for Arp 220, Mrk 231, IRAS 08572+3915, 
VV 114, and He 2--10.
Previously obtained data points of nearby LIRGs
\citep{ima04,ink06,in06}, starbursts, and Seyfert galaxies
\citep{koh05} are also plotted. 
As stated by \citet{ink06}, we mainly use the HCN(1--0)/HCO$^{+}$(1--0)
ratios (ordinate) in our discussions for the following reasons.
First, since the HCN(1--0) and HCO$^{+}$(1--0) lines are observed
simultaneously with the same array configuration of NMA/RAINBOW, 
their beam patterns are virtually identical. 
We can thus be confident that the same regions are probed with both
lines. 
Second, both HCN(1--0) and HCO$^{+}$(1--0) fluxes are measured at the same
time with the same receiver and same correlator unit so that the possible
{\it absolute} flux calibration uncertainty of interferometric data 
does not propagate to the uncertainty in the ratio, which is dominated by
statistical noise and fitting errors (see Figure 6). 
Therefore, the derivation of the HCN(1--0)/HCO$^{+}$(1--0)
brightness-temperature ratio is straightforward. 
Neither of the first and second arguments hold for the
HCN(1--0)/CO(1--0) brightness-temperature ratios in the abscissa of
Figure 8, which complicates their interpretation. 
LIRGs with luminous AGN signatures at other wavelengths (Arp
220W, Mrk 231, IRAS 08572+3915) tend to show high HCN(1--0)/HCO$^{+}$(1--0) 
brightness-temperature ratios as seen in AGN-dominated galaxies.
The ratio is also high in VV 114E-1, which is close to the putative AGN
location at VV 114E, despite a slight positional deviation. 
The Wolf-Rayet galaxy He 2--10 shows a very low HCN(1--0)/HCO$^{+}$(1--0) 
brightness-temperature ratio.
Some individual examples of the nearby Galactic high-mass-star-forming,
HII-region cores show high HCN(1--0)/HCO$^{+}$(1--0)  
brightness-temperature ratios \citep{tur77,pir99}. 
However, for external galaxies, the ensemble of HII regions, molecular
gas, and PDRs will be observed as a whole.
The low HCN(1--0)/HCO$^{+}$(1--0) brightness-temperature ratio in 
He 2--10 suggests that even an extreme starburst dominated by young hot
stars cannot easily produce a high HCN(1--0)/HCO$^{+}$(1--0)
brightness-temperature ratio when observed at a large physical scale. 
The high ratios observed in LIRGs are likely to have another
cause than the extreme starburst scenario.

\subsection{Interpretations of HCN(1--0)/HCO$^{+}$(1--0)
brightness-temperature ratios}

A natural explanation for the high HCN(1--0)/HCO$^{+}$(1--0) 
brightness-temperature ratios in buried AGN candidates is the 
increasing abundance of HCN due to strong X-ray radiation from the AGN, 
as predicted by some simple chemical calculations \citep{lep96,lin06}.
When both the HCN(1--0) and HCO$^{+}$(1--0) emission are optically
thin, the HCN(1--0)/HCO$^{+}$(1--0) brightness-temperature ratio will
increase linearly with increasing HCN abundance relative to
HCO$^{+}$. 
\citet{ink06} and Figure 9 show that even if the emission is
moderately optically thick, the increasing HCN abundance results in a
larger HCN(1--0)/HCO$^{+}$(1--0) brightness-temperature ratio 
(see also Knudsen et al. 2007).   
Since the HCN(1--0)/HCO$^{+}$(1--0) brightness-temperature ratios
in Figure 8 derived from our NMA/RAINBOW data are only toward the
cores where putative AGNs are expected to be present, with a minimum
contamination from extended starburst emission outside the beam size, 
the potential AGN's X-ray chemistry, if any, may better be reflected 
than single-dish telescope's measurements.

The critical density (n$_{\rm c}$) HCN(1--0) is a factor of $\sim$5
larger than that of HCO$^{+}$(1--0) ($\S$ 4.1). 
When the fraction of high density gas, above the HCN(1--0) critical
density, increases, the HCN(1--0)/HCO$^{+}$(1--0)
brightness-temperature ratio could be high compared to low density gas.
Since LIRGs have highly concentrated molecular gas in their nuclei
\citep{sam96} and the density may be larger than the normal starburst
galaxies studied by \citet{koh05}, the large HCN(1--0)/HCO$^{+}$(1--0)
brightness-temperature ratios could originate simply in the increased
molecular gas density.
However, in this scenario, the HCN(1--0)/CO(1--0) brightness-temperature
ratios should be correspondingly high (see $\S$4.1).
Although HCN(1--0)/CO(1--0) brightness-temperature ratios in the
abscissa of Figure 8 exhibit some ambiguities ($\S$ 5.1), no such
expected trend is observed in Figure 8 (LIRGs are not distributed in the
right side), which does not support this high gas-density scenario.
The expected high gas density in the nuclei of LIRGs and non-high
HCN(1--0)/CO(1--0) brightness-temperature ratios can be reconciled
if the higher average gas density in certain areas results from a larger
volume-filling factor of the clump (Figure 9), rather than the higher
density of each clump.

\citet{gar06} found a very large HCN(5--4)/HCO$^{+}$(5--4)
brightness-temperature ratio in a luminous AGN with strong mid-infrared
12 $\mu$m continuum emission (APM 08279+5255), and suggested that
infrared radiative pumping, rather 
than collisional excitation, can produce a high HCN/HCO$^{+}$
brightness-temperature ratio. 
In general, an AGN is a more luminous mid-infrared 12 $\mu$m continuum
emitter than a starburst, because 12 $\mu$m hot dust emission is
stronger in the former. 
Hence, the infrared pumping may work more effectively in LIRGs with
luminous buried AGNs than those without, which could also 
explain the high HCN(1--0)/HCO$^{+}$(1--0) brightness-temperature ratios
seen in luminous buried AGN candidates. 

\citet{pap07} proposed that the HCN(1--0)/HCO$^{+}$(1--0)
brightness-temperature ratio could increase with increasing turbulence
of molecular gas. 
In this model, it is expected that the ratio is higher with larger line
widths of HCN(1--0) and HCO$^{+}$(1--0). 
Figure 10 compares the line width with the ratio, but we see no clear
positive trend.

\citet{gra06} measured the HCN(1--0) and HCO$^{+}$(1--0) fluxes of LIRGs
using a single-dish telescope with $>$25-arcsec beam size, and 
found a positive correlation between the HCN(1--0)/HCO$^{+}$(1--0) 
brightness-temperature ratio and infrared luminosity. 
Figure 11 shows the HCN(1--0)/HCO$^{+}$(1--0) brightness-temperature
ratio only for the {\it core} emission inside 1.5--10 arcsec beam sizes
as a function of infrared luminosity from the whole galactic region. 
Among LIRGs plotted in Figure 11, the two LIRGs dominated by a single
main nucleus, Mrk 231 and Mrk 273, were plotted in the figure of
\citet{gra06}. 
The two LIRGs with multiple main nuclei, Arp 220 and Arp 299, 
were also studied by \citet{gra06}, but our interferometric data 
resolved the nuclei, providing essentially new plots.
Other LIRGs in Figure 11 are new. 
A positive correlation is also recognizable here with a different sample.  
If our AGN's X-ray interpretation for the origin of high
HCN(1--0)/HCO$^{+}$(1--0) brightness-temperature ratios is correct, then
an increasing importance of AGNs at the cores of LIRGs with
higher infrared luminosities is suggested. 
Investigating the spatial variation of the ratios using a better
spatial-resolution map (i.e., separating a true compact core and
surrounding extended regions) may help to clarify the origin
of the ratios. 
Detailed theoretical calculations that solve chemical reaction networks
in a self-consistent way and realistically account for the clumpy structure of 
molecular gas are also desirable for interpreting our
observational results.

\section{Summary}

We presented the results of millimeter interferometric HCN(1--0) and
HCO$^{+}$(1--0) observations of nearby LIRGs using the NMA/RAINBOW array.
When combined with other LIRGs we have previously observed, 
ours is the largest interferometric HCN(1--0) and HCO$^{+}$(1--0) survey
of LIRGs. 
Most of the observed LIRGs are selected based on the presence of 
luminous AGN signatures at other wavelengths. 
From the interferometric data, we extracted the HCN(1--0) and
HCO$^{+}$(1--0) flux at the core positions, where the putative AGNs
are expected to be present. 
For observationally investigating the possible chemical
effects of the putative strongly X-ray-emitting AGN to the surrounding
interstellar gas, our data set is much more advantageous than single-dish
telescope's data because the contamination from extended starburst
emission can be reduced. 
We derived the HCN(1--0)/HCO$^{+}$(1--0) brightness-temperature ratios 
of LIRGs and compared them to the ratios found in AGNs and starburst
galaxies. 

We reached the following main conclusions. 
\begin{enumerate}

\item LIRGs with detected luminous AGN signatures through observations
      at other wavelengths generally show high HCN(1--0)/HCO$^{+}$(1--0)
      brightness-temperature ratios and are distributed in 
      the region occupied by AGNs.
\item When combined with HCN(1--0)/CO(1--0) brightness-temperature
      ratios and the line widths of HCN(1--0) and HCO$^{+}$(1--0),
      the high HCN(1--0)/HCO$^{+}$(1--0) brightness-temperature ratios
      observed in these LIRGs are naturally explained by an HCN abundance 
      enhancement as suggested by some simple theoretical models. 
\item We found a positive correlation between the
      HCN(1--0)/HCO$^{+}$(1--0) brightness-temperature ratios only
      toward the LIRGs' cores and infrared luminosities of LIRGs, suggesting
      that AGN activity is more important in galaxies with higher
      infrared luminosities.  
\end{enumerate}

\acknowledgments

We are grateful to the NRO staff for their support during our
NMA/RAINBOW observing runs, and T. Pyo, R. Potter, S. Harasawa 
for their assistance during our Subaru observing runs. 
We thank J. Gracia-Carpio and S. Ishizuki for valuable discussions about
molecular gas, and the anonymous referee for his/her useful comments. 
M.I. is supported by Grants-in-Aid for Scientific Research (16740117 and
19740109). 
Y.T. is financially supported by the Japan Society for the Promotion of
Science (JSPS) for Young Scientists. 
K.K. is supported by the MEXT Grant-in-Aid for Scientific Research on
Priority Areas (15071202).
NRO is a branch of the National Astronomical Observatory, National
Institutes of Natural Sciences, Japan.
This study utilized the SIMBAD database, operated at CDS,
Strasbourg, France, and the NASA/IPAC Extragalactic Database
(NED) operated by the Jet Propulsion Laboratory, California
Institute of Technology, under contract with the National Aeronautics
and Space Administration.

\clearpage

\begin{deluxetable}{lcccrrcl}
\tabletypesize{\scriptsize}
\tablecaption{Detailed information about the observed LIRGs and the
Wolf-Rayet galaxy He 2--10
\label{tab1}}
\tablewidth{0pt}
\tablehead{
\colhead{Object} & \colhead{Redshift}   & 
\colhead{$f_{\rm 12}$}  & \colhead{$f_{\rm 25}$}  & 
\colhead{$f_{\rm 60}$}  & \colhead{$f_{\rm 100}$}  & 
\colhead{log $L_{\rm IR}$ (log $L_{\rm IR}$/$L_{\odot}$)} &
\colhead{Far-infrared} \\ 
\colhead{} & \colhead{}   & \colhead{(Jy)} & \colhead{(Jy)}  & \colhead{(Jy)} 
& \colhead{(Jy)}  & \colhead{(ergs s$^{-1}$)} & \colhead{Color} \\
\colhead{(1)} & \colhead{(2)} & \colhead{(3)} & \colhead{(4)} & 
\colhead{(5)} & \colhead{(6)} & \colhead{(7)} & \colhead{(8)}
}
\startdata
Arp 220 & 0.018 & 0.48 & 7.92 & 103.33 & 112.40 & 45.7 (12.1) & 0.08 (cool) \\
Mrk 231 & 0.042 & 1.87 & 8.66 & 31.99 & 30.29 & 46.1 (12.5) & 0.27 (warm) \\ 
IRAS 08572+3915 & 0.058 & 0.32 & 1.70 & 7.43 & 4.59 & 45.7 (12.1) & 0.23
(warm) \\ 
VV 114 (IC 1623)  & 0.020 & 0.68 & 3.57 & 22.58 & 30.37 & 45.2 (11.6) &
0.16 (cool) \\ 
He 2--10 & 0.003 & 1.09 & 6.51 & 24.08 & 26.40 & 43.5 (9.9)
\tablenotemark{a} & 0.27 (warm) \\ 
\enddata

\tablecomments{
Column (1): Object name.
Column (2): Redshift.
Columns (3)--(6): f$_{12}$, f$_{25}$, f$_{60}$, and f$_{100}$ are
{\it IRAS FSC} (or {\it PSC}) 
fluxes at 12, 25, 60, and 100 $\mu$m, respectively.
Column (7): Decimal logarithm of the infrared (8$-$1000 $\mu$m) luminosity
in ergs s$^{-1}$ calculated as follows:
$L_{\rm IR} = 2.1 \times 10^{39} \times$ D(Mpc)$^{2}$
$\times$ (13.48 $\times$ $f_{12}$ + 5.16 $\times$ $f_{25}$ +
$2.58 \times f_{60} + f_{100}$) ergs s$^{-1}$
\citep{sam96}.
The values in parentheses are the decimal logarithms of the infrared
luminosities in units of solar luminosities.
Column (8): {\it IRAS} 25 $\mu$m to 60 $\mu$m flux ratio
(f$_{25}$/f$_{60}$). LIRGs with f$_{25}$/f$_{60}$ $<$ ($>$) 0.2 are
classified as cool (warm) \citep{san88b}.
}

\tablenotetext{a}{
Distance of 10.5 Mpc \citep{tul88} is adopted.}

\end{deluxetable}

\begin{deluxetable}{lclccll}
\tabletypesize{\scriptsize}
\tablecaption{Observing log\label{tab2}}  
\tablewidth{0pt}
\tablehead{
\colhead{Object} & \colhead{Array} & \colhead{Observing Date} &
\colhead{Central} & \colhead{Phase} & \colhead{Band-pass} & \colhead{Flux} \\ 
\colhead{} & \colhead{configuration} & \colhead{(UT)} &
\colhead{frequency} & \colhead{calibrator} & \colhead{calibrator} &
\colhead{calibrator} \\  
\colhead{(1)} & \colhead{(2)} & \colhead{(3)} & \colhead{(4)} &
\colhead{(5)} & \colhead{(6)} & \colhead{(7)}
}
\startdata
Arp 220  & RAINBOW & 2006 Jan 26, 28, 30 & 87.34 & 1538+149 & 3C 273 &
1741$-$038\\ 
         & AB & 2007 Jan 22, 23, 24, 25 & & & & \\ \hline
Mrk 231  & RAINBOW & 2006 Jan 25, 26, 28 & 85.31 & 1418+546 & 3C 273 &
1741$-$038\\  
\hline
IRAS 08572+3915 & RAINBOW & 2005 Jan 28 & 84.04 & 0923+392 & 3C 84, 3C 273
& 3C 84\\ 
                & AB & 2006 Feb 12 & & & & \\
                & C  & 2005 Mar 13, 14, 25, Dec 11, 13 & & & & \\
                & D  & 2006 Mar 4, 5 & & & & \\ \hline
VV 114 & AB & 2007 Jan 29, 30, Feb 6, 7, 8 & 87.16 & 0048$-$097 &
3C 84, 3C 454.3 & Uranus \\
      & C  & 2007 Mar 29, 30 & & & & \\
      & D  & 2006 Nov 28, 29, 30, Dec 1 & & & & \\ 
      &    & 2007 Apr 26, 27 & & & & \\ \hline
He 2--10 & AB & 2007 Feb 6 & 88.64 & 0834$-$201 & 3C 84, 3C 279 & 3C 84 \\
       & C  & 2007 Mar 28, 29, 30 & & & & \\
       & D  & 2006 Dec 13, 2007 Apr 26, 27& & & & \\
\enddata

\tablecomments{
Column (1): Object name.
Column (2): NMA array configuration.
Column (3): Observing date in UT.
Column (4): Central frequency used for the observations.
Column (5): Object name used as a phase calibrator.
Column (6): Object name used as a band-pass calibrator. 
Column (7): Object name used as a flux calibrator. 
}

\end{deluxetable}

\begin{deluxetable}{lrcc}
\tablecaption{Parameters of final NMA/RAINBOW map \label{tab3}}  
\tablewidth{0pt}
\tablehead{
\colhead{Object} & \colhead{On source} & \colhead{Beam size} &
\colhead{Position angle of} \\  
\colhead{} & \colhead{integration (hr)} & \colhead{(arcsec $\times$ arcsec)} &
\colhead{the beam ($^{\circ}$)} \\
\colhead{(1)} & \colhead{(2)} & \colhead{(3)} & \colhead{(4)}
}
\startdata
Arp 220 & 24 & 1.8 $\times$ 1.5 & $-$55.1 \\
Mrk 231 & 4  & 2.1 $\times$ 1.8 & $-$34.8 \\
IRAS 08572+3915 & 33 & 4.7 $\times$ 3.5 & $-$50.7 \\
VV 114  & 18 & 7.5 $\times$ 5.5 & $-$13.2\\
He 2--10 & 13 & 10.8 $\times$ 5.5 & $-$10.1 \\
\enddata

\tablecomments{
Col.(1): Object name.
Col.(2): Net on-source integration time in hours.
Col.(3): Beam size in arcsec $\times$ arcsec.
Col.(4): Position angle of the beam pattern.
It is 0$^{\circ}$ for the north-south direction, and increases 
counterclockwise on the sky plane.
}

\end{deluxetable}

\begin{deluxetable}{llccclccc}
\tabletypesize{\scriptsize}
\tablecaption{Subaru IRCS observing log \label{tab4}}
\tablewidth{0pt}
\tablehead{
\colhead{Object} & 
\colhead{Date} & 
\colhead{Integration} & 
\colhead{Slit width}  & 
\colhead{P.A.}  & 
\multicolumn{4}{c}{Standard Stars} \\
\colhead{} & 
\colhead{(UT)} & 
\colhead{(Min)} &
\colhead{($''$)} &
\colhead{($^{\circ}$)} &
\colhead{Name} &  
\colhead{$L$-mag} &  
\colhead{Type} &  
\colhead{$T_{\rm eff}$ (K)}  \\
\colhead{(1)} & \colhead{(2)} & \colhead{(3)} & \colhead{(4)}
& \colhead{(5)} & \colhead{(6)}  & \colhead{(7)} & \colhead{(8)} &
\colhead{(9)}  
}
\startdata
Arp 220 & 2006 July 17    & 24 & 0.6 & 87 & HR 5728 & 4.5 & G3V & 5800 \\  
        & 2007 January 14 & 20 & 0.6 & 87 & HR 5728 & 4.5 & G3V & 5800 \\  
Mrk 231 & 2007 January 14 & 16 & 0.6 & 90 & HR 4767 & 4.8 & F8-G0V & 6000 \\
IRAS 08572+3915NW & 2007 January 15 & 24 & 0.9 & 90 & HR 3451 & 5.0 & F7V &
6240 \\  
VV 114E & 2006 July 19 & 24 & 0.6 & 60 & HR 210 & 4.0 & G3V & 5800 \\  
\hline  
\enddata

\tablecomments{Column (1): Object name.
Col. (2): Observing date in UT.
Col. (3): Net on-source integration time in min.
Col. (4): Slit width in arcsec.
Col. (5): Position angle of the slit.
0$^{\circ}$ corresponds to the north-south direction.
Position angle increases counterclockwise on the sky plane.
Col. (6): Standard star name.
Col. (7): Adopted $L$-band magnitude.
Col. (8): Stellar spectral type.
Col. (9): Effective temperature.}

\end{deluxetable}

\begin{deluxetable}{lcc}
\tablecaption{Continuum emission \label{tab5}}  
\tablewidth{0pt}
\tablehead{
\colhead{Object} & \colhead{Flux} & \colhead{Frequency} \\ 
\colhead{} & \colhead{(mJy)} & \colhead{(GHz)} 
}
\startdata
Arp 220 & 30 & $\sim$87 \\
Mrk 231 & 35 & $\sim$85 \\
IRAS 08572+3915NW & $<$3 ($<3\sigma$) & $\sim$84 \\
VV 114  & $<$4 ($<3\sigma$) & $\sim$87 \\
He 2--10 & $<$6 ($<3\sigma$) & $\sim$88.5 \\
\enddata
\end{deluxetable}

\begin{deluxetable}{lccccc}
\tabletypesize{\scriptsize}
\tablecaption{Gaussian fitting parameters of HCN(1--0) and
HCO$^{+}$(1--0) emission lines \label{tab6}}   
\tablewidth{0pt}
\tablehead{
\colhead{Object} & \multicolumn{3}{c}{LSR velocity} &
\multicolumn{2}{c}{FWHM}  \\   
\colhead{} & \multicolumn{3}{c}{(km s$^{-1}$)} & 
\multicolumn{2}{c}{(km s$^{-1}$)} \\ 
\colhead{} & \colhead{HCN(1--0)} & \colhead{HCO$^{+}$(1--0)} &
\colhead{CO(1--0) or CO(2--1)} & \colhead{HCN(1--0)} & 
\colhead{HCO$^{+}$(1--0)}  \\
\colhead{(1)} & \colhead{(2)} & \colhead{(3)} & \colhead{(4)} &
\colhead{(5)} & \colhead{(6)} 
}
\startdata
Arp 220E & 5640 & 5680 & 5650 & 240 & 190 \\
Arp 220W & 5310 & 5290 & 5340 & 320 & 270  \\
Mrk 231 & 12660 & 12690 & 12650--12660 & 260 & 290 \\
IRAS 08572+3915NW & 17110 \tablenotemark{a} + 17540 \tablenotemark{a} &
17250 \tablenotemark{a} + 17900 \tablenotemark{a} & 17490 &
420 \tablenotemark{a} + 300 \tablenotemark{a} & 
320 \tablenotemark{a} + 180 \tablenotemark{a} \\
VV 114E-1 & 6090 & \nodata & \nodata & 170 & \nodata \\
VV 114E-2 & 6070 & 6060 & \nodata & 190 & 260 \\
VV 114-3  & 6060 & 6030 & \nodata & 290 & 360 \\
VV 114-4  & 5910 & 5950 & \nodata & 250 & 310 \\
\enddata

\tablecomments{
Col.(1): Object name.
Col.(2): LSR velocity \{v$_{\rm opt}$ $\equiv$ 
($\frac{\nu_0}{\nu}$ $-$ 1) $\times$ c\} of the
HCN(1--0) emission peak.
Col.(3): LSR velocity of the HCO$^{+}$(1--0) emission peak.
Col.(4): LSR velocity of the CO(1--0) or CO(2--1) emission peak, 
taken from the literature \citep{bry96,dow98,eva02}.
Col.(5): Line width of the HCN(1--0) emission at FWHM.
Col.(6): Line width of the HCO$^{+}$(1--0) emission at FWHM.
}

\tablenotetext{a}{A double Gaussian fit.}

\end{deluxetable}

\begin{deluxetable}{lccccc}
\tabletypesize{\scriptsize}
\tablecaption{Flux of HCN(1--0), HCO$^{+}$(1--0), and CO(1--0) emission
at each peak position \label{tab7}}    
\tablewidth{0pt}
\tablehead{
\colhead{Nucleus} & \colhead{HCN(1--0)} & \colhead{HCO$^{+}$(1--0)} &
\colhead{HCN(1--0)/HCO$^{+}$(1--0)} & \colhead{CO(1--0)} &
\colhead{Reference} \\    
\colhead{} & \colhead{(Jy km s$^{-1}$)} & \colhead{(Jy km s$^{-1}$)} & 
\colhead{} & \colhead{(Jy km s$^{-1}$)} & \colhead{} \\ 
\colhead{(1)} & \colhead{(2)} & \colhead{(3)} & \colhead{(4)} & 
\colhead{(5)} & \colhead{(6)}
}
\startdata
Arp 220E & 11.9 & 5.3 & 2.3 & 30 \tablenotemark{a} & A \\
Arp 220W & 17.8 & 6.6 & 2.7 & 47 \tablenotemark{a} & A \\
Mrk 231  & 10.7 & 6.3 & 1.7 & 68 & B \\
IRAS 08572+3915NW & 2.3 (1.4 + 0.9) \tablenotemark{b} & 1.3 (0.7 + 0.6)
\tablenotemark{b} & 1.8 & 11.8 & C \\ 
VV 114E-1 & 2.7 & $<$1.7 & $>$1.6 & $<$70 \tablenotemark{c} & D \\
VV 114E-2 & 1.9 & 4.0 & 0.5 & \nodata & \nodata \\
VV 114-3  & 3.2 & 5.5 & 0.6 & \nodata & \nodata \\
VV 114-4  & 1.3 & 6.0 & 0.2 & \nodata & \nodata \\
He 2--10  & 6.1 \tablenotemark{d} & 11 \tablenotemark{d} & $<$0.6
\tablenotemark{e} & 30 & E \\
\enddata

\tablecomments{
Col.(1): Object name.
Col.(2): Integrated HCN(1--0) intensity at an emission peak.
Col.(3): Integrated HCO$^{+}$(1--0) intensity at an emission peak.
Col.(4): HCN(1--0)/HCO$^{+}$(1--0) ratio in brightness temperature 
($\propto$ $\lambda^{2}$ $\times$ flux density) at an emission peak.
The ratio is not affected by possible absolute flux uncertainties in
the NMA/RAINBOW data (see $\S$5.1).
Col.(5): Integrated CO(1--0) intensity at an emission peak taken from the
literature. 
Col.(6): References for interferometric CO(1--0) flux measurements.
(A): \citet{sak99}. (B): \citet{dow98}. (C): \citet{eva02}
(D): \citet{ion04}. (E): \citet{kob95}. 
}

\tablenotetext{a}{
CO(2--1) emission is detected at individual cores, but 
CO(1--0) emission is not \citep{dow98}.
We estimate the CO(1--0) flux {\it at the core} from the CO(2--1)
flux \citep{sak99}, using the $\nu^{2}$ scaling \citep{rie06}, based 
on the assumption that emission is thermalized and that  
both CO(1--0) and CO(2--1) have the same brightness temperatures.}

\tablenotetext{b}{Blue (former) and red (latter) components from the
double Gaussian fits (see $\S$4.1).}

\tablenotetext{c}{CO(1--0) flux is estimated from CO(3--2)
flux, by assuming the $\nu^{2}$ scaling.}

\tablenotetext{d}{
Flux is derived from the peak contour of the integrated intensity map.  
Possible continuum emission is not subtracted, because
it is difficult to estimate in a reliable way due to the large scatter
of the spectrum.}

\tablenotetext{e}{The ratio is an upper limit because 
possible continuum emission is not subtracted and HCN(1--0) emission is 
weaker than HCO$^{+}$(1--0) emission.}

\end{deluxetable}

\begin{deluxetable}{lccccc}
\tabletypesize{\scriptsize}
\tablecaption{Total flux of HCN(1--0) and HCO$^{+}$(1--0) emission for
sources that show spatially-extended components \label{tab8}}   
\tablewidth{0pt}
\tablehead{
\colhead{Nucleus} & \colhead{HCN(1--0)} & \colhead{HCO$^{+}$(1--0)}  \\   
\colhead{} & \colhead{(Jy km s$^{-1}$)} & \colhead{(Jy km s$^{-1}$)} \\ 
\colhead{(1)} & \colhead{(2)} & \colhead{(3)} }
\startdata
Arp 220 & 34 & 18 \\
VV 114  & 7 & 19 \\
He 2--10 & 6.5 & 11 \\
\enddata

\tablecomments{
Col.(1): Object name.
Col.(2): Total HCN(1--0) flux.
Col.(3): Total HCO$^{+}$(1--0) flux.
}

\end{deluxetable}

\begin{deluxetable}{llcl}
\tabletypesize{\scriptsize}
\tablecaption{Comparison of HCN(1--0) and HCO$^{+}$(1--0) fluxes between our
measurements and those in the literature \label{tab9}}   
\tablewidth{0pt}
\tablehead{
\colhead{Nucleus} & \colhead{Line} & \colhead{Flux} & \colhead{Reference} \\   
\colhead{(1)} & \colhead{(2)} & \colhead{(3)}  & \colhead{(4)} 
}
\startdata
Arp 220 & HCN(1--0) & 34\tablenotemark{a} & This work \\
        &     & 45 & \citet{eva06} \\ 
        &     & 37 & \citet{sol92} \\ 
        &     & 35\tablenotemark{a} & \citet{rad91}  \\ 
        & HCO$^{+}$(1--0) & 18\tablenotemark{a} & This work \\
        &           & 23 & \citet{gra06} \\ \hline
Mrk 231 & HCN(1--0) & 11\tablenotemark{a} & This work \\ 
        &     & 14 & \citet{sol92} \\ 
        & HCO$^{+}$(1--0) & 6\tablenotemark{a} & This work \\
        &           & 8 & \citet{gra06} \\
\enddata

\tablecomments{
Col.(1): Object name.
Col.(2): HCN(1--0) or HCO$^{+}$(1--0) line.
Col.(3): Flux in [Jy km s$^{-1}$].
Col.(4): Reference.
}

\tablenotetext{a}{Measurements are made based on interferometric data.}

\end{deluxetable}

\begin{deluxetable}{lcccc}
\tabletypesize{\scriptsize}
\tablecaption{Strength of the 3.3 $\mu$m PAH emission feature \label{tab10}}   
\tablewidth{0pt}
\tablehead{
\colhead{Object} & \colhead{f$_{3.3 \rm PAH}$} & \colhead{L$_{3.3 \rm PAH}$} &
\colhead{L$_{3.3 \rm PAH}$/L$_{\rm IR}$}  & 
\colhead{rest EW$_{3.3 \rm PAH}$} \\   
\colhead{} & 
\colhead{($\times$10$^{-14}$ ergs s$^{-1}$ cm$^{-2}$)} & 
\colhead{($\times$10$^{40}$ergs s$^{-1}$)}  & 
\colhead{($\times$10$^{-3}$)} & 
\colhead{(nm)} \\
\colhead{(1)} & \colhead{(2)} & \colhead{(3)} & \colhead{(4)} & \colhead{(5)} }
\startdata
Arp 220E   & 3.5  & 2.2 & \nodata & 85 \\
Arp 220W   & 7.5  & 4.8 & \nodata & 95 \\
Arp 220E+W & 11.0 & 6.9 & 0.01 & 95 \\
Mrk 231     & 9.5  & 33.5 & 0.03 & 1.5 \\
IRAS 08572+3915NW  & $<$4.0 & $<$30.0 & $<$0.05 & $<$3 \\
VV 114E    & 3.0  & 2.5 & \nodata & 20 \\
VV 114E$_{\rm SW}$ & 2.5  & 2.0 & \nodata & 110 \\
\enddata

\tablecomments{
Col.(1): Object name. 
Col.(2): Observed flux of the 3.3 $\mu$m PAH emission. 
Col.(3): Observed luminosity of the 3.3 $\mu$m PAH emission.  
Col.(4): Observed 3.3 $\mu$m PAH-to-infrared luminosity ratio in units
         of 10$^{-3}$, a typical value for a modestly-obscured 
         (A$_{\rm V}$ $<$ 15 mag) normal starburst \citep{mou90,ima02}.    
         For Arp 220E, W and VV 114E, E$_{\rm SW}$, no values are shown,
         because the fraction of {\it IRAS}-measured infrared fluxes, coming
         from individual nuclei, is unknown.
Col.(5): Rest-frame equivalent width of the 3.3 $\mu$m PAH emission.  
         Normal starbursts have EW$_{\rm 3.3PAH}$ $\sim$ 100 nm \citep{imd00}.
}

\end{deluxetable}

\clearpage

\begin{figure}
\includegraphics[angle=0,scale=.66]{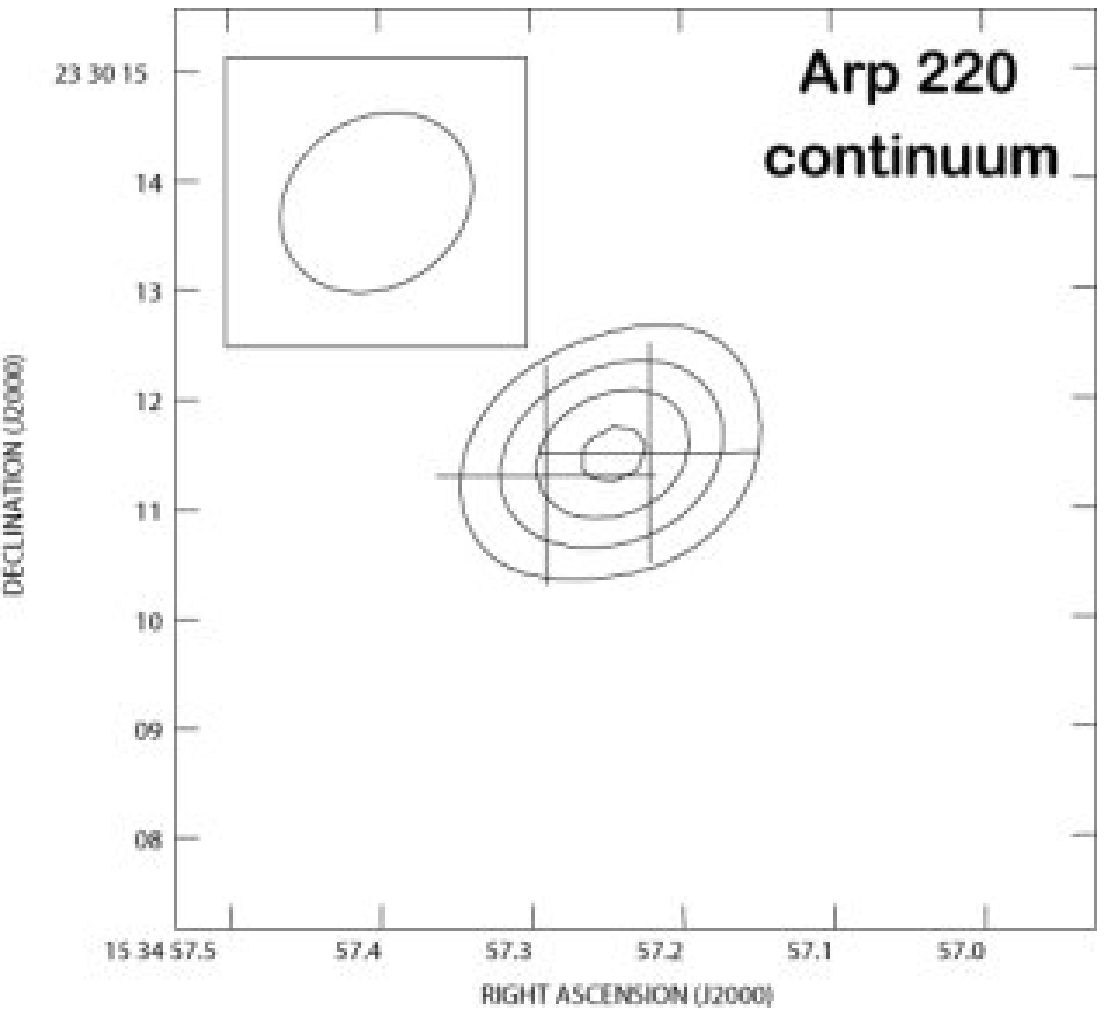} 
\includegraphics[angle=0,scale=.66]{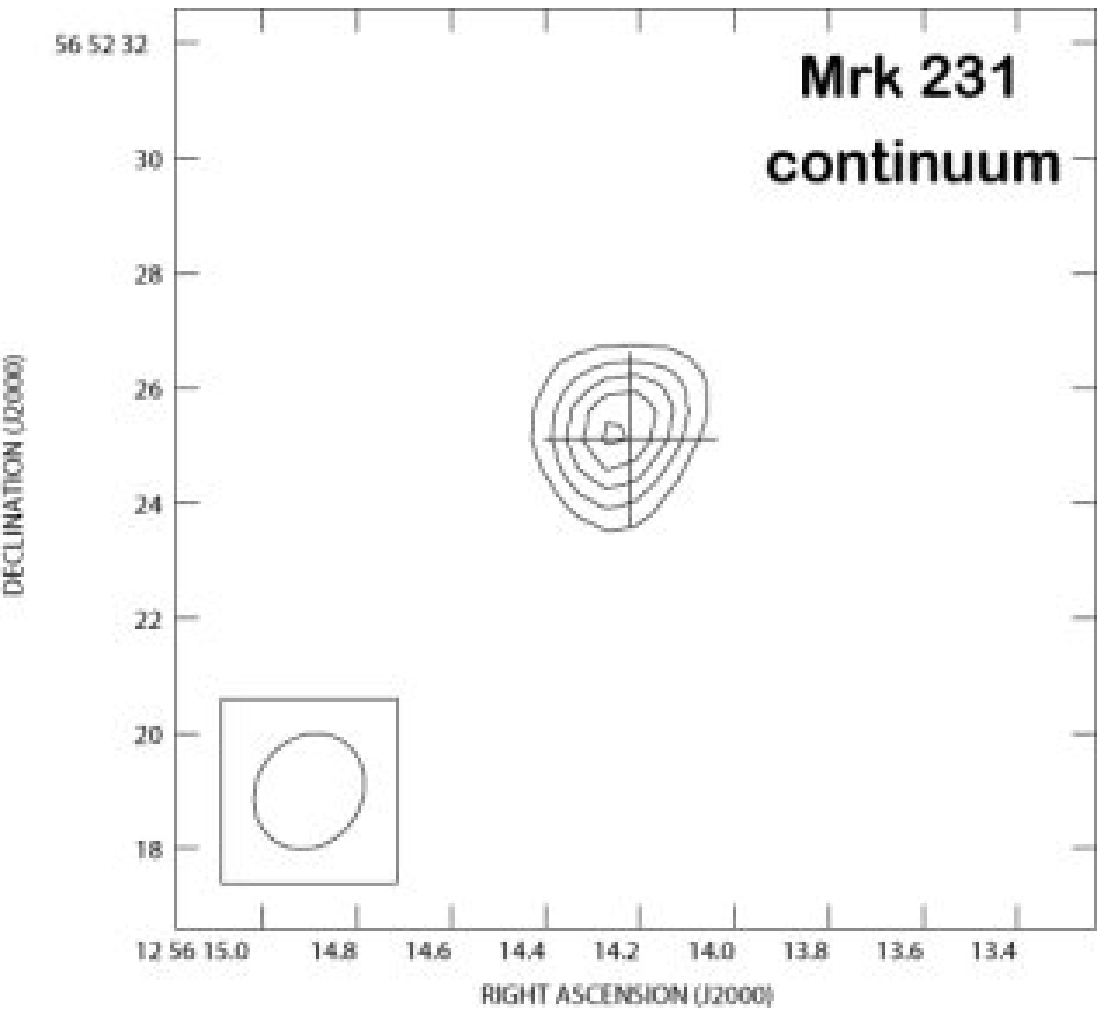} 
\caption{
Continuum maps of Arp 220 at $\sim$87 GHz and Mrk 231 at $\sim$85 GHz.
The crosses show the coordinates of main nuclei.
The coordinates in J2000 are
(15$^{h}$34$^{m}$57.22$^{s}$, $+$23$^{\circ}$30$'$11$\farcs$5) for 
Arp 220W, 
(15$^{h}$34$^{m}$57.29$^{s}$, $+$23$^{\circ}$30$'$11$\farcs$3) for
Arp 220E, and
(12$^{h}$56$^{m}$14.22$^{s}$, $+$56$^{\circ}$52$'$25$\farcs$1) for
Mrk 231.
These nuclear coordinates are adopted from CO(2--1) peaks in millimeter
interferometric maps \citep{dow98}. 
The contours are 3 $\times$ (3, 5, 7, 9) mJy beam$^{-1}$ for Arp 220,
3 $\times$ (3, 5, 7, 9, 11) mJy beam$^{-1}$ for Mrk 231.
}
\end{figure}

\clearpage

\begin{figure}
\includegraphics[angle=0,scale=.66]{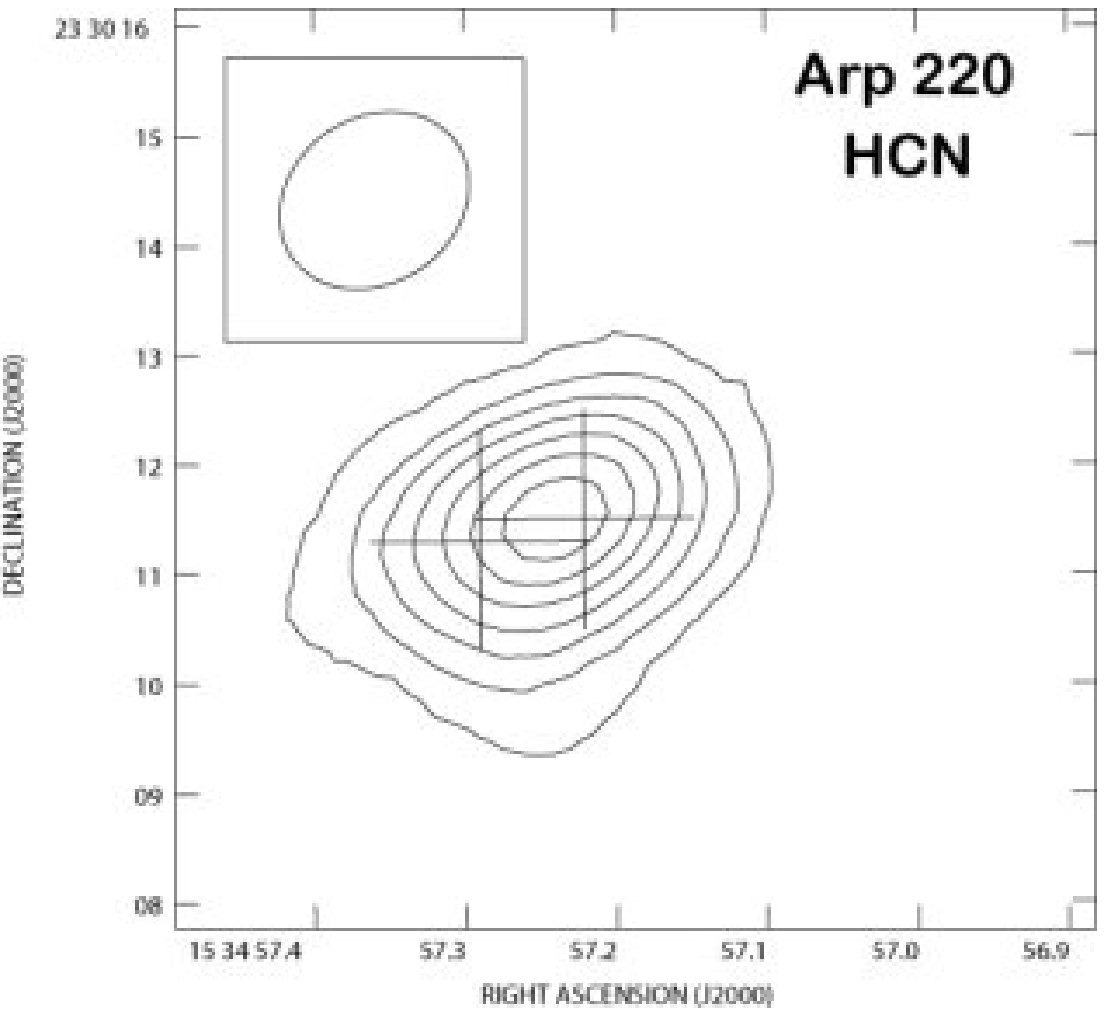} 
\includegraphics[angle=0,scale=.66]{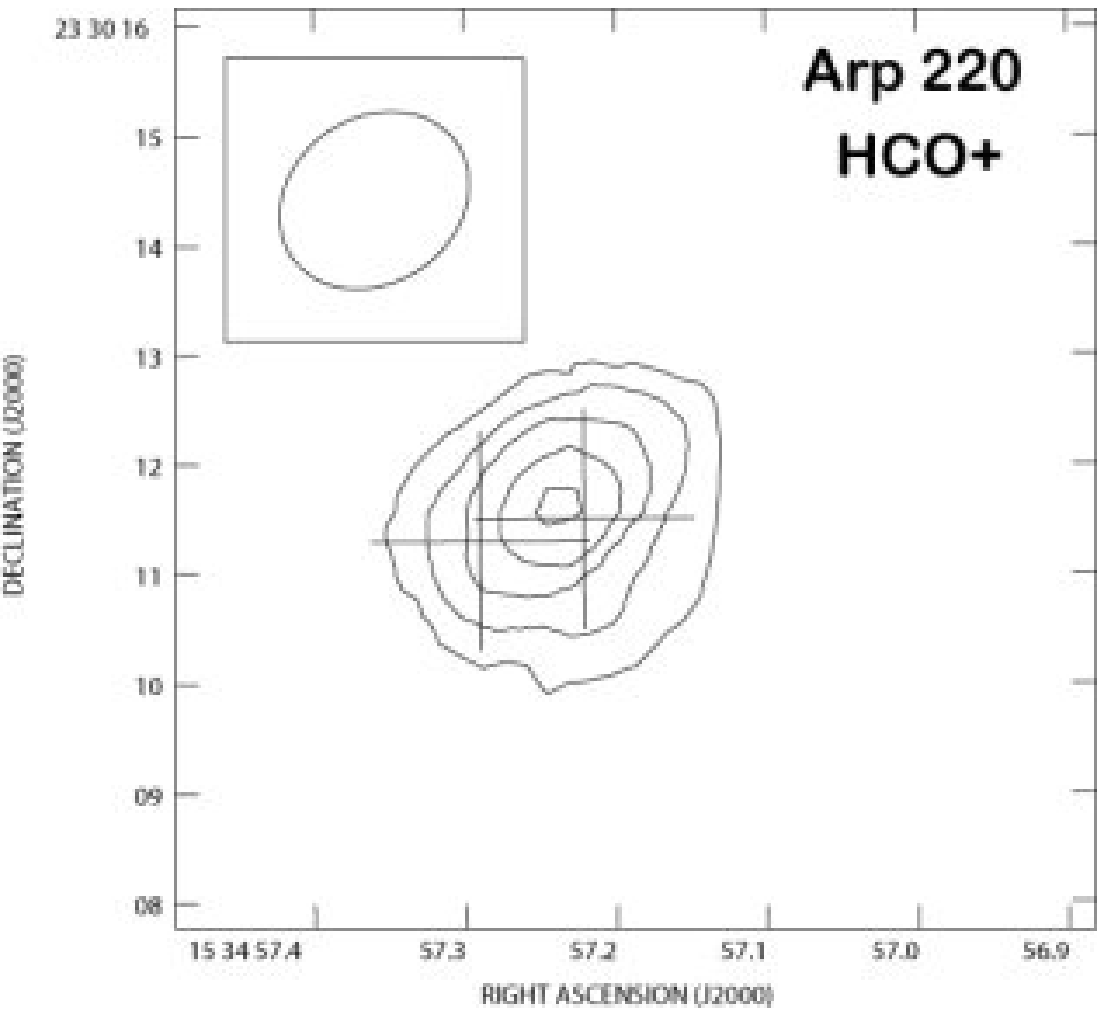} 
\includegraphics[angle=0,scale=.66]{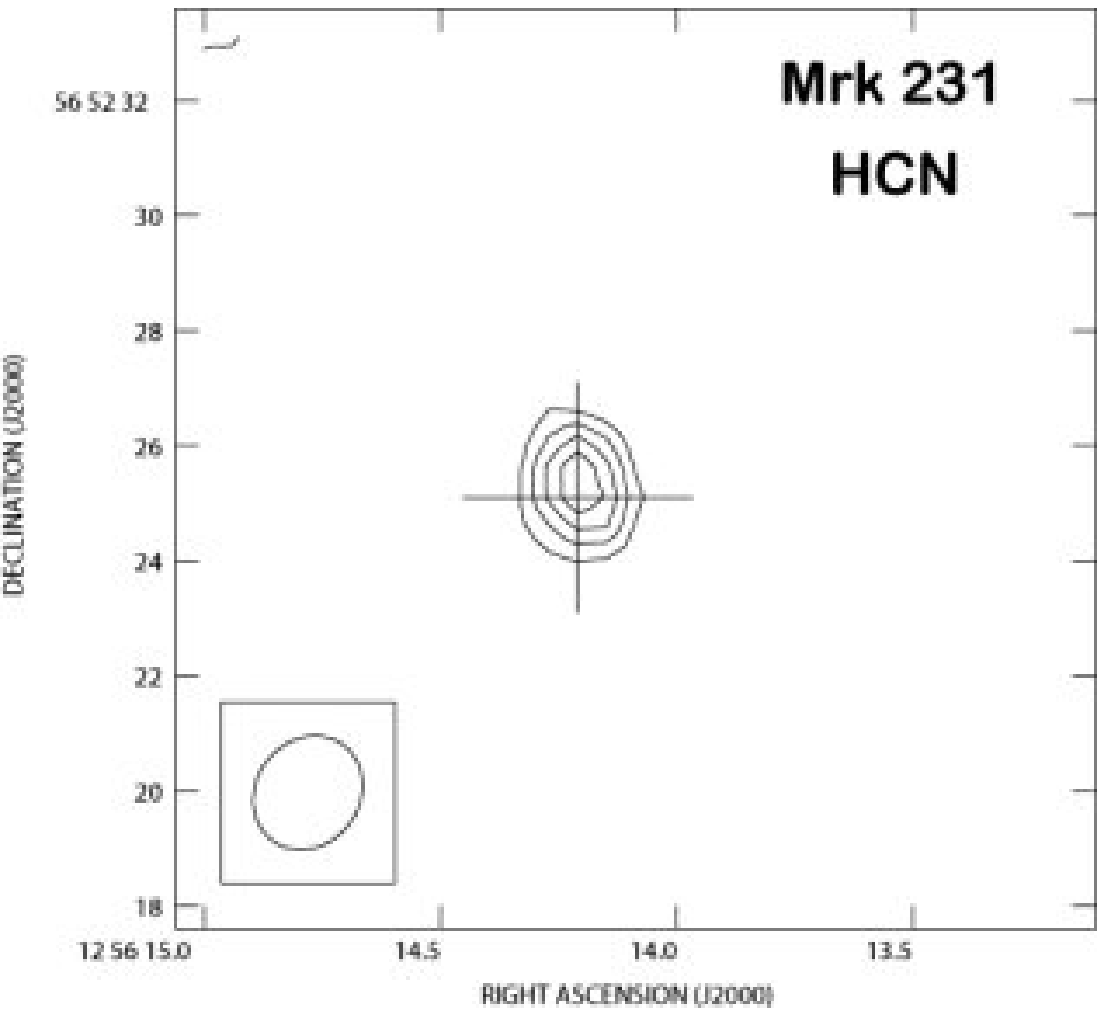}
\includegraphics[angle=0,scale=.66]{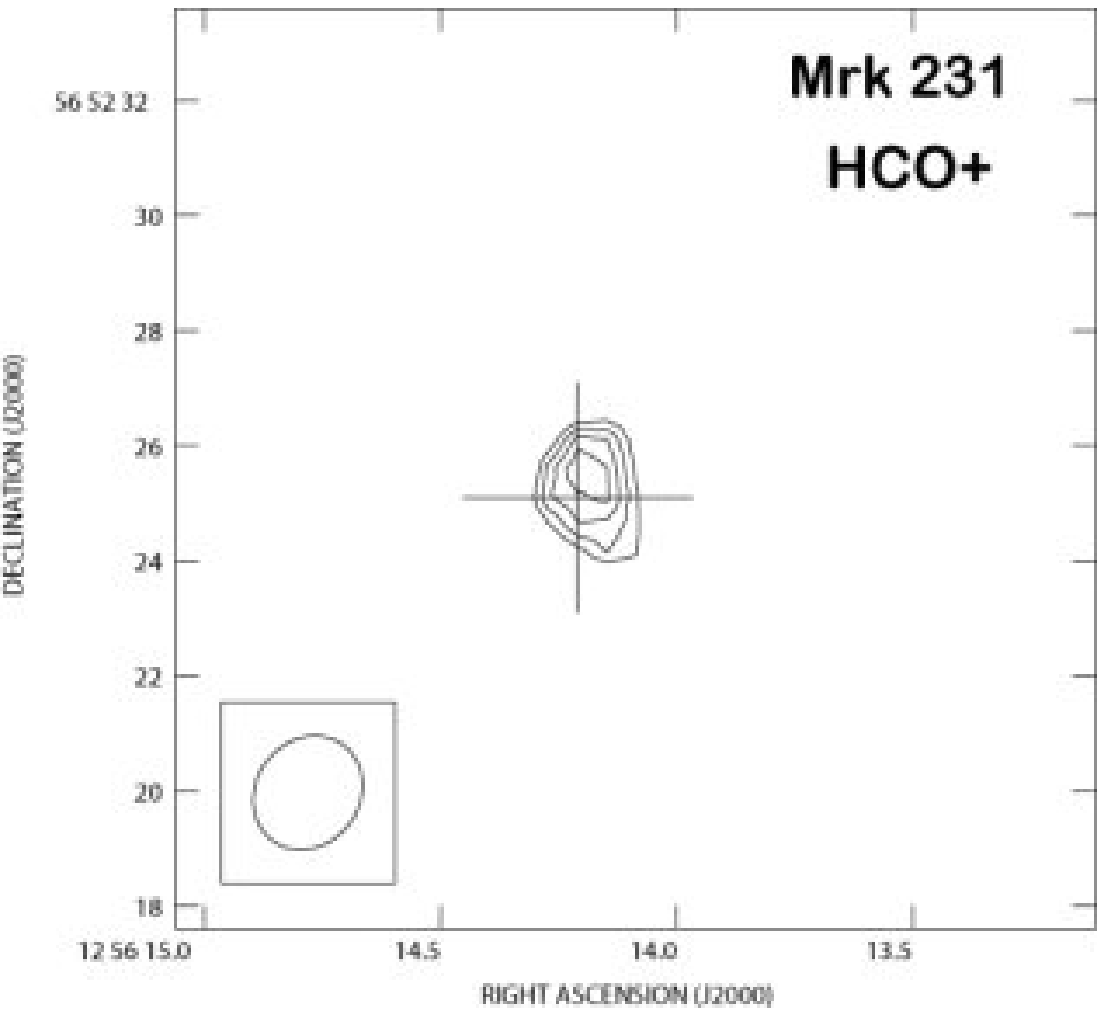} 
\includegraphics[angle=0,scale=.66]{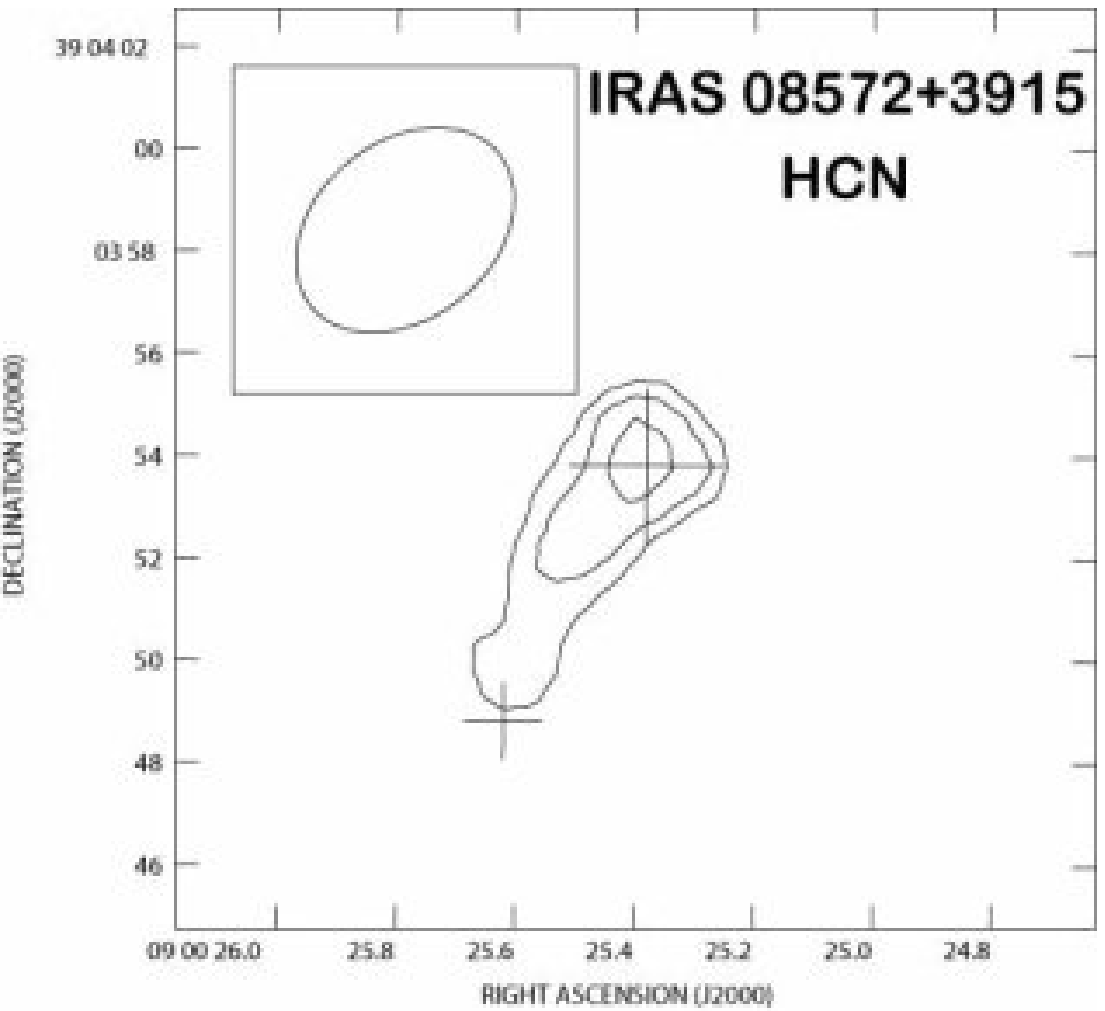}
\hspace{1.2cm}
\includegraphics[angle=0,scale=.66]{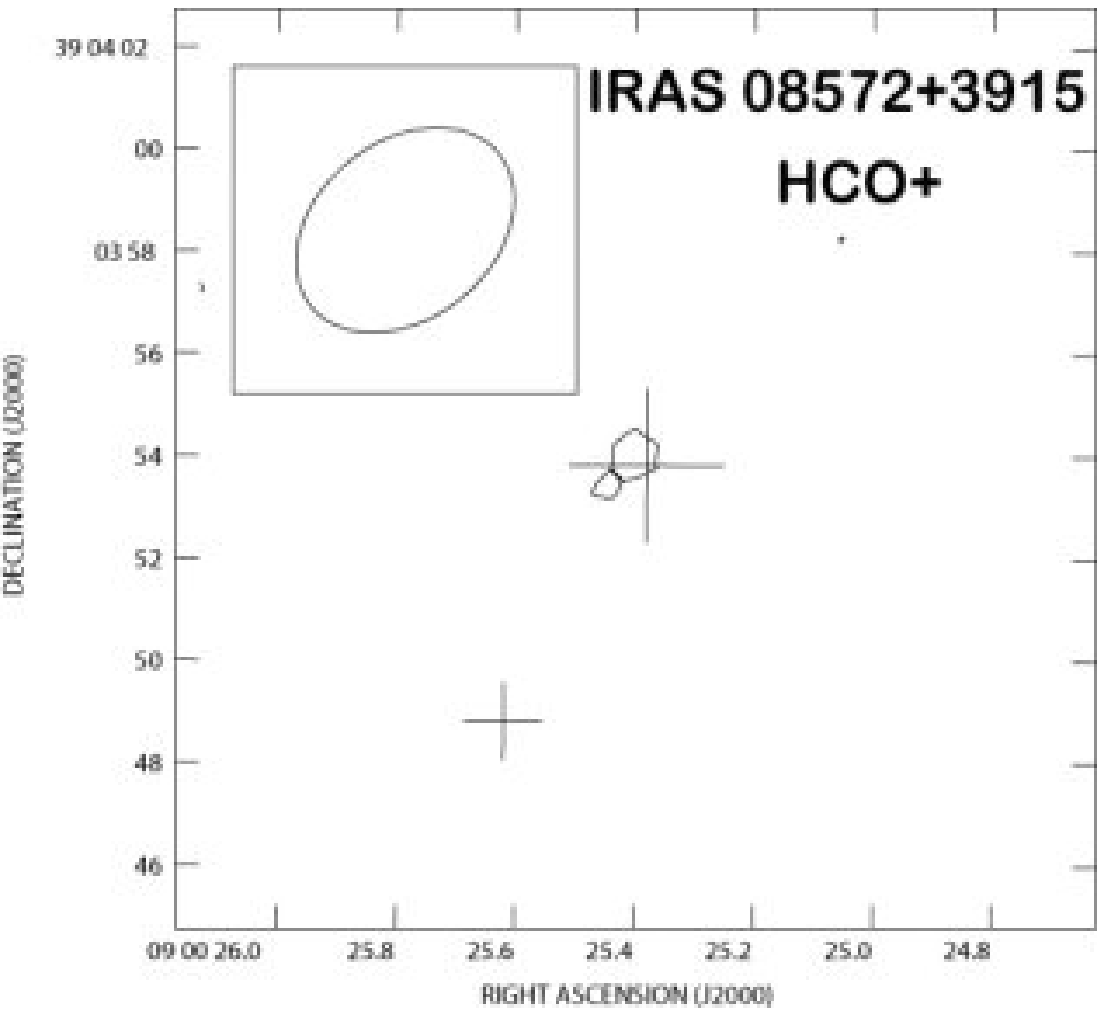}
\end{figure}

\clearpage
\begin{figure}
\includegraphics[angle=0,scale=.66]{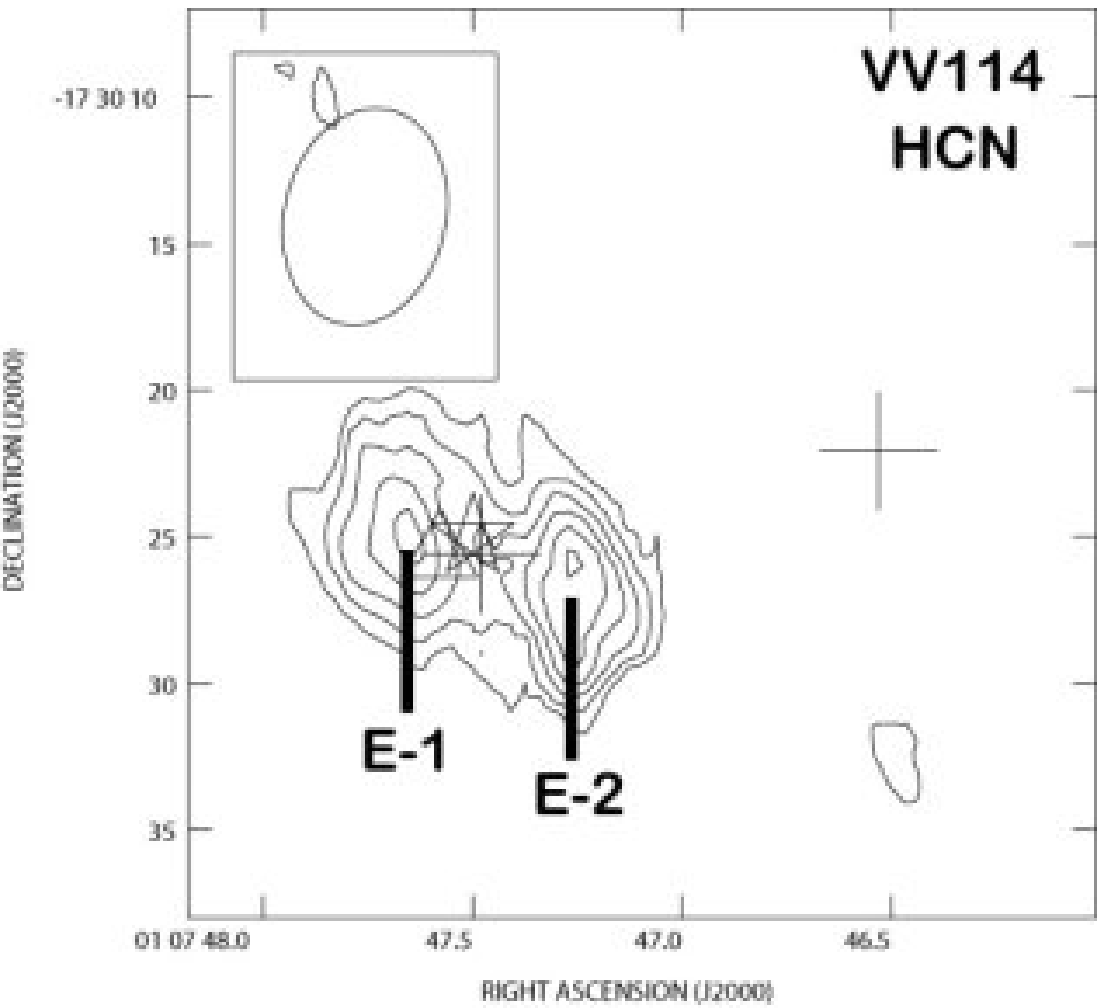}
\includegraphics[angle=0,scale=.66]{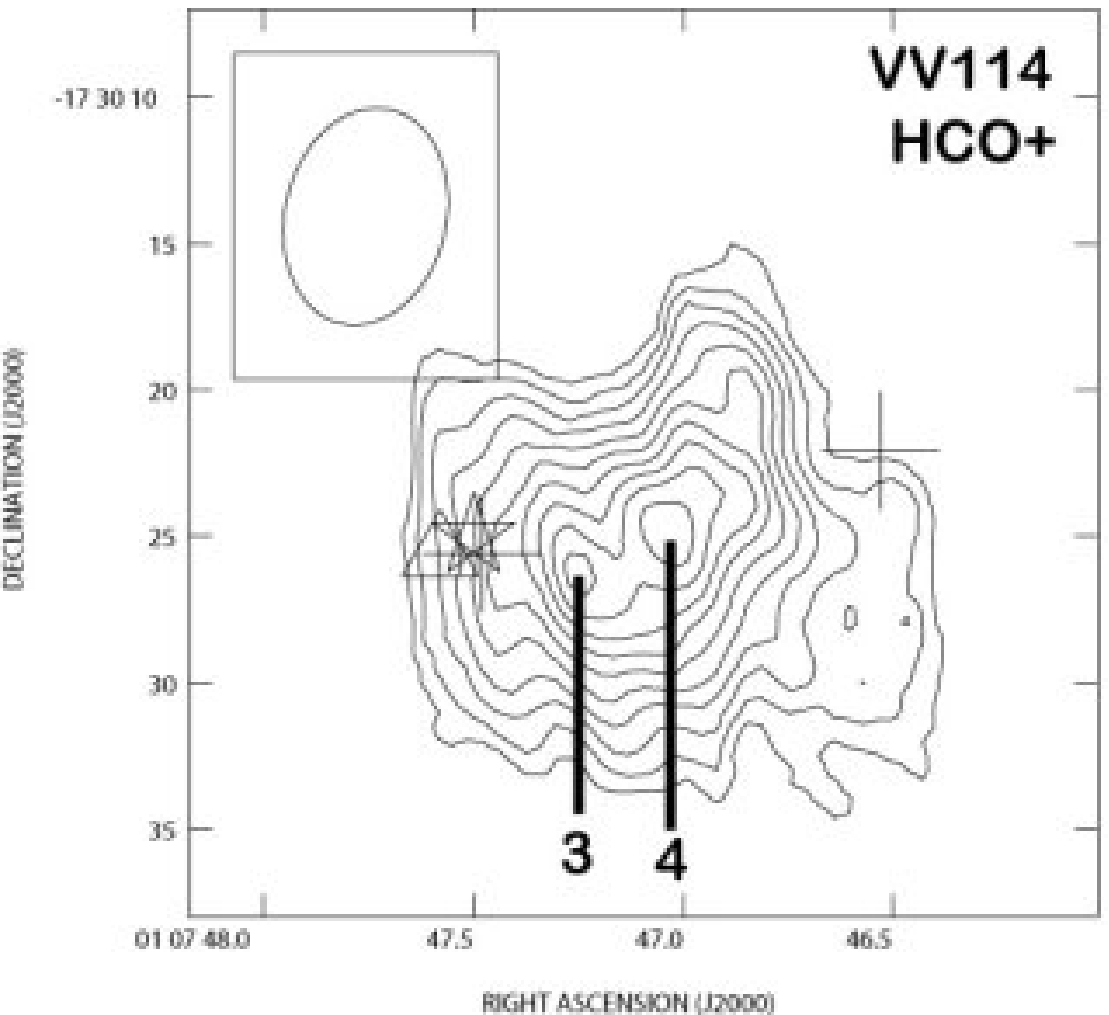}
\includegraphics[angle=0,scale=.66]{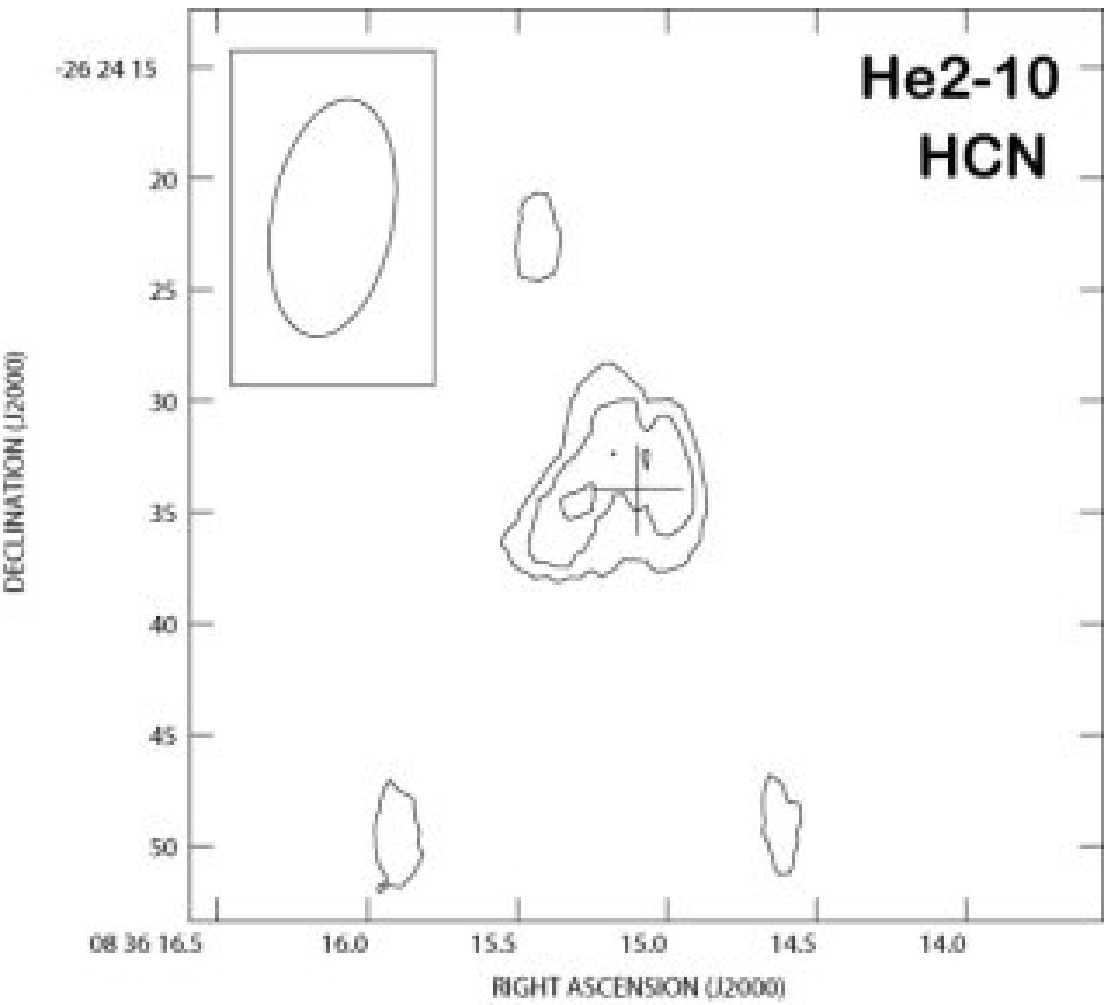} 
\hspace{1.2cm}
\includegraphics[angle=0,scale=.66]{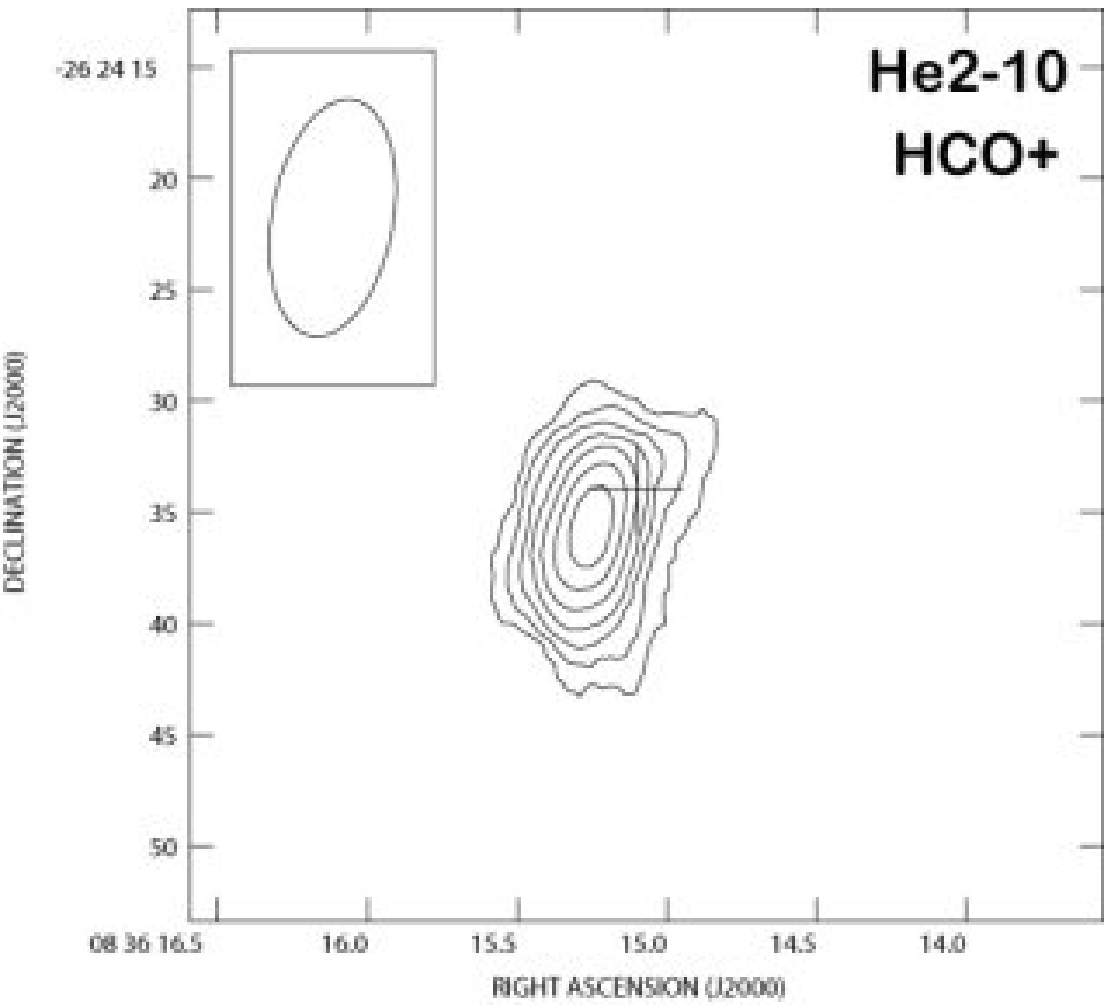} 
\caption{\small 
{\it Left}: HCN(1--0) emission map.
The contours are 0.85 $\times$ (3, 7, 11, 15, 19, 23, 27) Jy km s$^{-1}$
beam$^{-1}$ for Arp 220, 
0.76 $\times$ (4, 6, 8, 10) Jy km s$^{-1}$ beam$^{-1}$ for Mrk 231, 
0.6 $\times$ (3, 4, 5) Jy km s$^{-1}$ beam$^{-1}$ for IRAS 08572+3915, 
0.5 $\times$ (3, 4, 5, 6, 7, 8) Jy km s$^{-1}$ beam$^{-1}$ for VV 114,
and 
1.2 $\times$ (3, 4, 5) Jy km s$^{-1}$ beam$^{-1}$ for He 2--10.
For IRAS 08572+3915, the two crosses correspond to the NW 
(9$^{h}$0$^{m}$25.38$^{s}$, $+$39$^{\circ}$3$'$53$\farcs$8) and SE 
(9$^{h}$0$^{m}$25.62$^{s}$, $+$39$^{\circ}$3$'$48$\farcs$8) peaks 
in J2000 \citep{eva02}.
For VV 114, the two crosses are near-infrared peaks, 
VV 114E (1$^{h}$7$^{m}$47.48$^{s}$, $-$17$^{\circ}$30$'$25$\farcs$6) and
VV 114W (1$^{h}$7$^{m}$46.53$^{s}$, $-$17$^{\circ}$30$'$22$\farcs$1)  
in J2000, whose coordinates are converted from B1950 \citep{kno94},
using NED.  
The open star is CO(3--2) peak 
(1$^{h}$7$^{m}$47.5$^{s}$, $-$17$^{\circ}$30$'$25$\farcs$0) in J2000
\citep{ion04}, and the open triangle is a radio 3.55 cm (8.44 GHz) peak 
(1$^{h}$7$^{m}$47.58$^{s}$, $-$17$^{\circ}$30$'$25$\farcs$6) in J2000, 
converted from B1950 \citep{con91} using NED.
E-1 and E-2 correspond to the two emission peaks in the HCN(1--0) map. 
For He 2--10, the cross is CO (1--0) peak 
(8$^{h}$36$^{m}$15.1$^{s}$, $-$26$^{\circ}$24$'$34$\farcs$0) in J2000,
converted from B1950 \citep{kob95} using NED. 
{\it Right}: HCO$^{+}$(1--0) emission map.
The contours are 0.85 $\times$ (3, 5, 7, 9, 11) Jy km s$^{-1}$
beam$^{-1}$ for Arp 220, 
0.76 $\times$ (3, 4, 5, 6) Jy km s$^{-1}$ beam$^{-1}$ for Mrk 231,
0.6 $\times$ 3 Jy km s$^{-1}$ beam$^{-1}$ for IRAS 08572+3915, 
0.5 $\times$ (3, 4, 5, 6, 7, 8, 9, 10, 11, 12) Jy km s$^{-1}$
beam$^{-1}$ for VV 114, and 
1.2 $\times$ (3, 4, 5, 6, 7, 8, 9) Jy km s$^{-1}$ beam$^{-1}$ for He 2--10.
For VV 114, the positions 3 and 4 correspond to the two emission peaks
in the HCO$^{+}$(1--0) map. 
For IRAS 08572+3915 and He 2--10, possible continuum emission is not
subtracted from both the HCN(1--0) and HCO$^{+}$(1--0) maps (see $\S$4.1). 
}
\end{figure}

\clearpage

\begin{figure}
\includegraphics[angle=0,scale=.6]{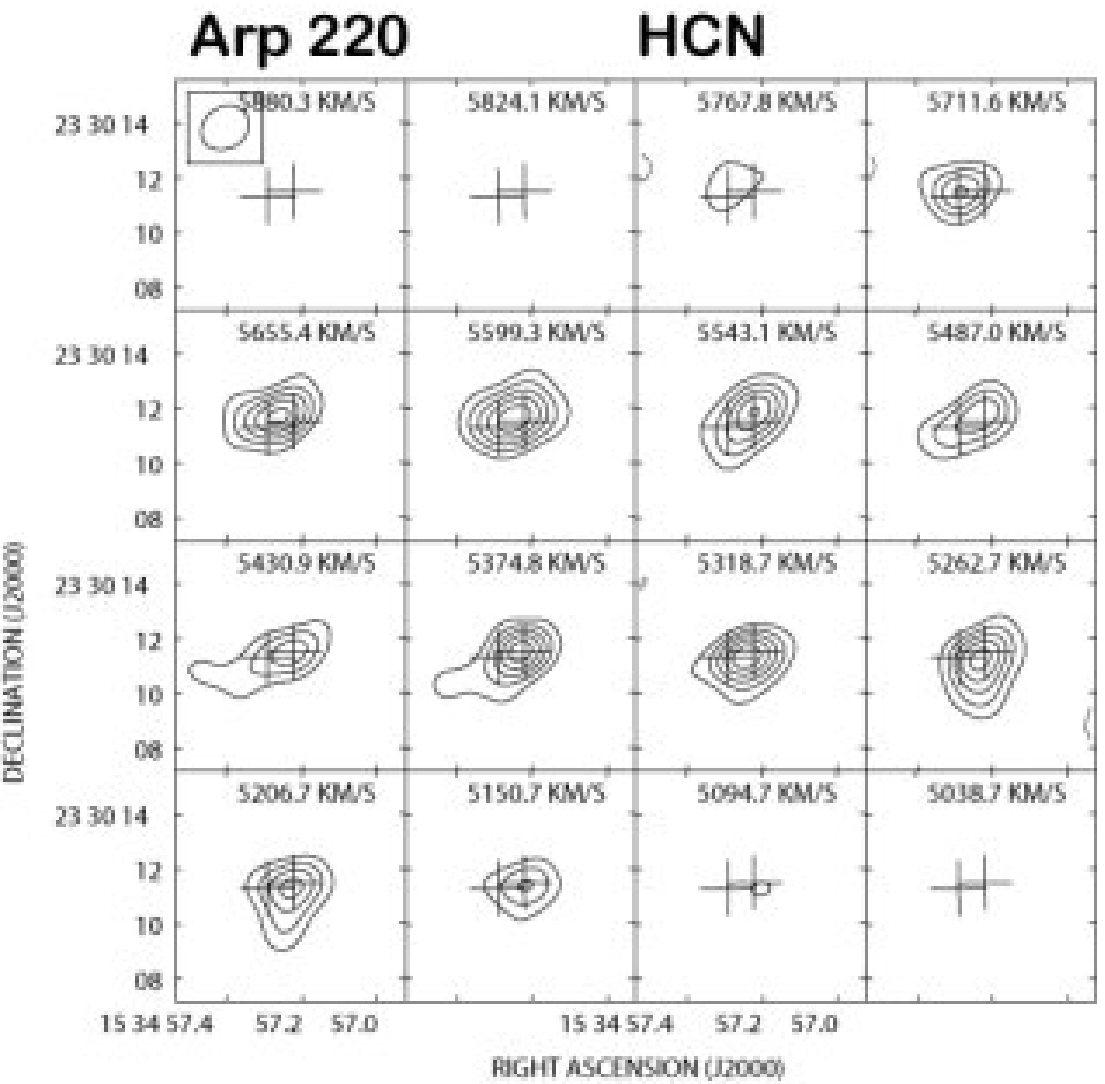}
\includegraphics[angle=0,scale=.6]{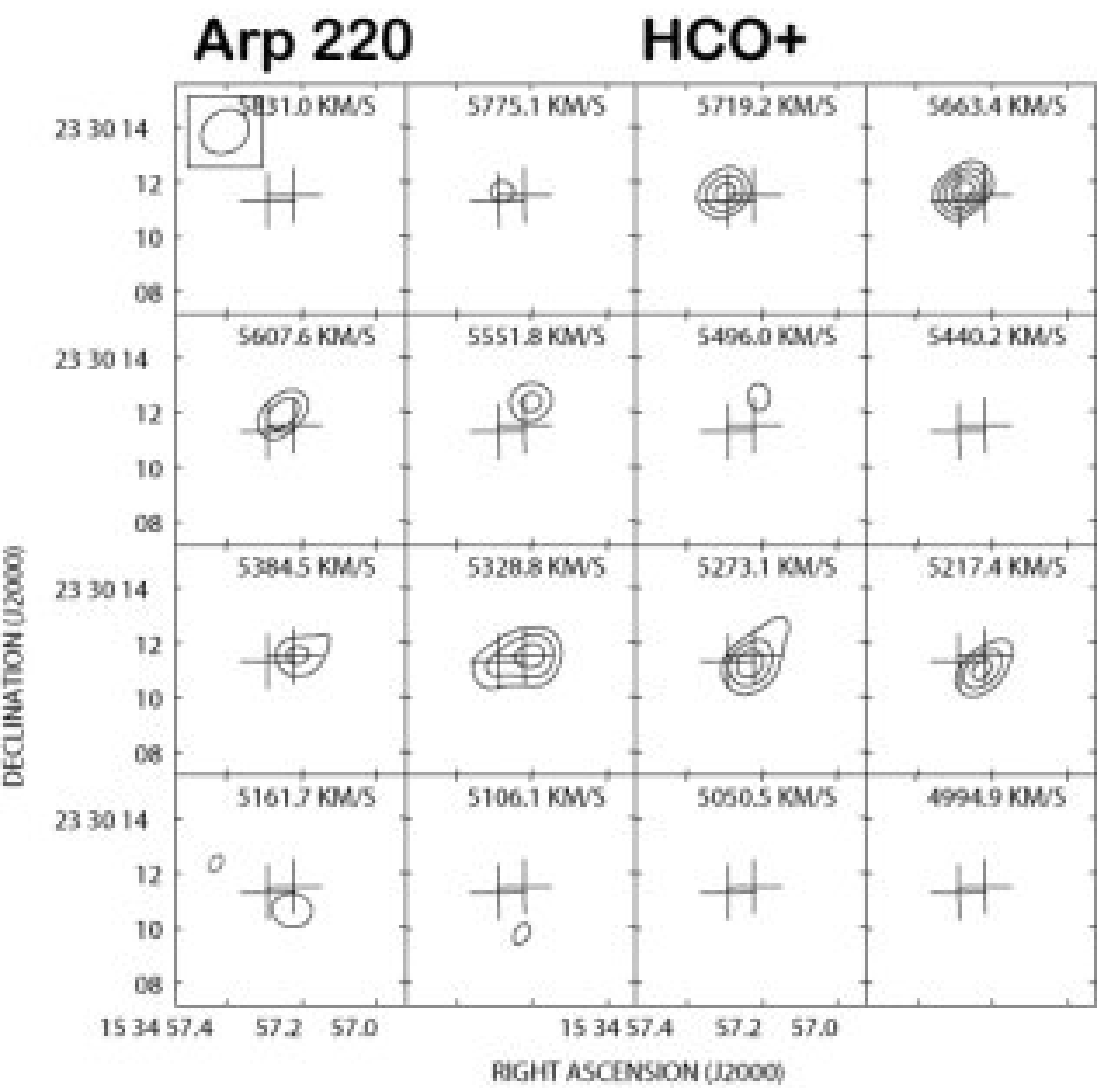}
\caption{
Channel maps of HCN(1--0) and HCO$^{+}$(1--0) emission for Arp 220.
{\it Left}: HCN(1--0) emission.
The contours are 4 $\times$ ($-$3, 3, 5, 7, 9, 11) mJy beam$^{-1}$.
The r.m.s. noise level is $\sim$4 mJy beam$^{-1}$.
{\it Right}: HCO$^{+}$(1--0) emission.
The contours are 4 $\times$ ($-$3, 3, 4, 5, 6) mJy beam$^{-1}$.
The r.m.s. noise level is $\sim$4 mJy beam$^{-1}$.
}
\end{figure}

\clearpage

\begin{figure}
\includegraphics[angle=0,scale=.6]{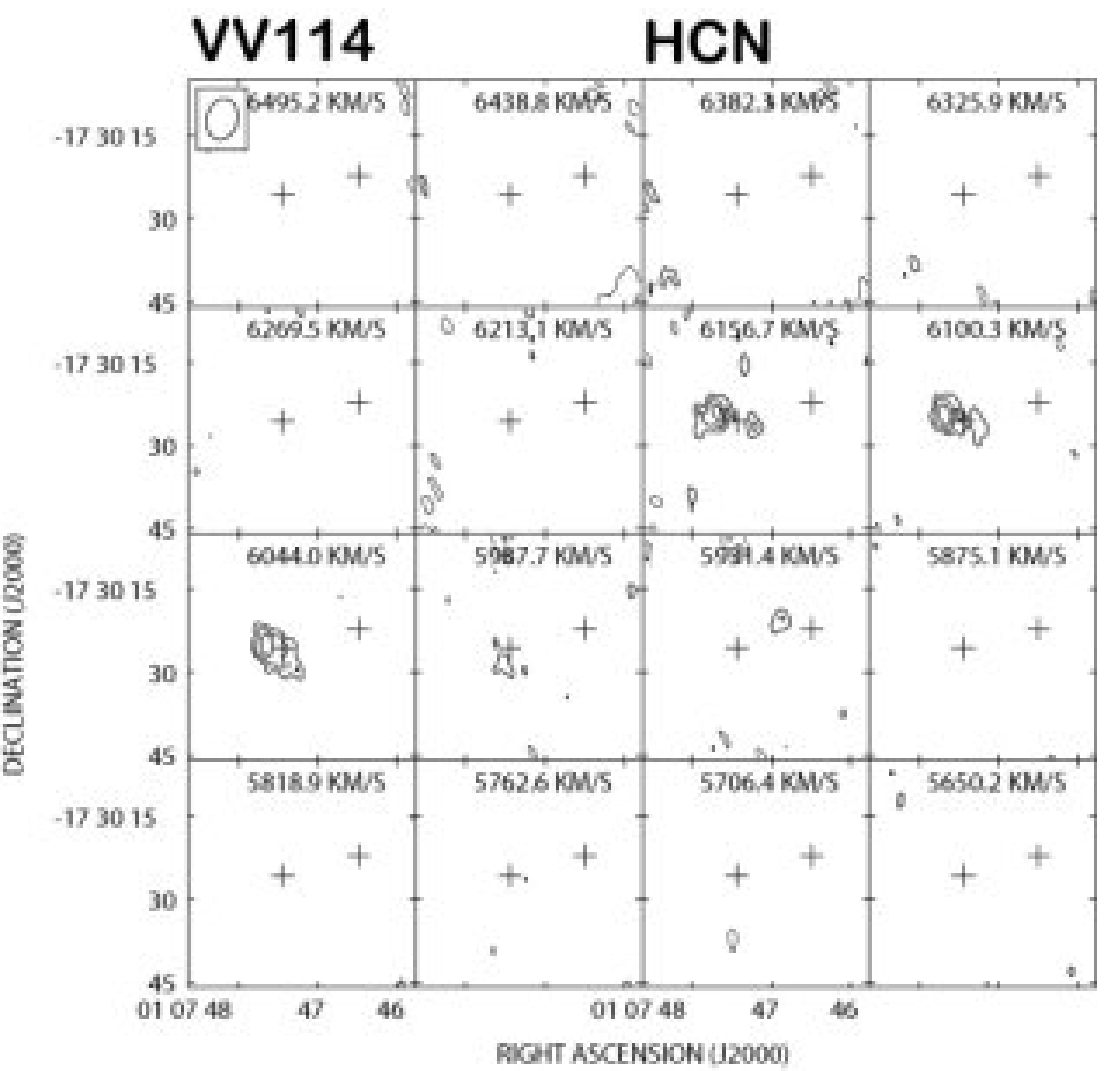}
\includegraphics[angle=0,scale=.6]{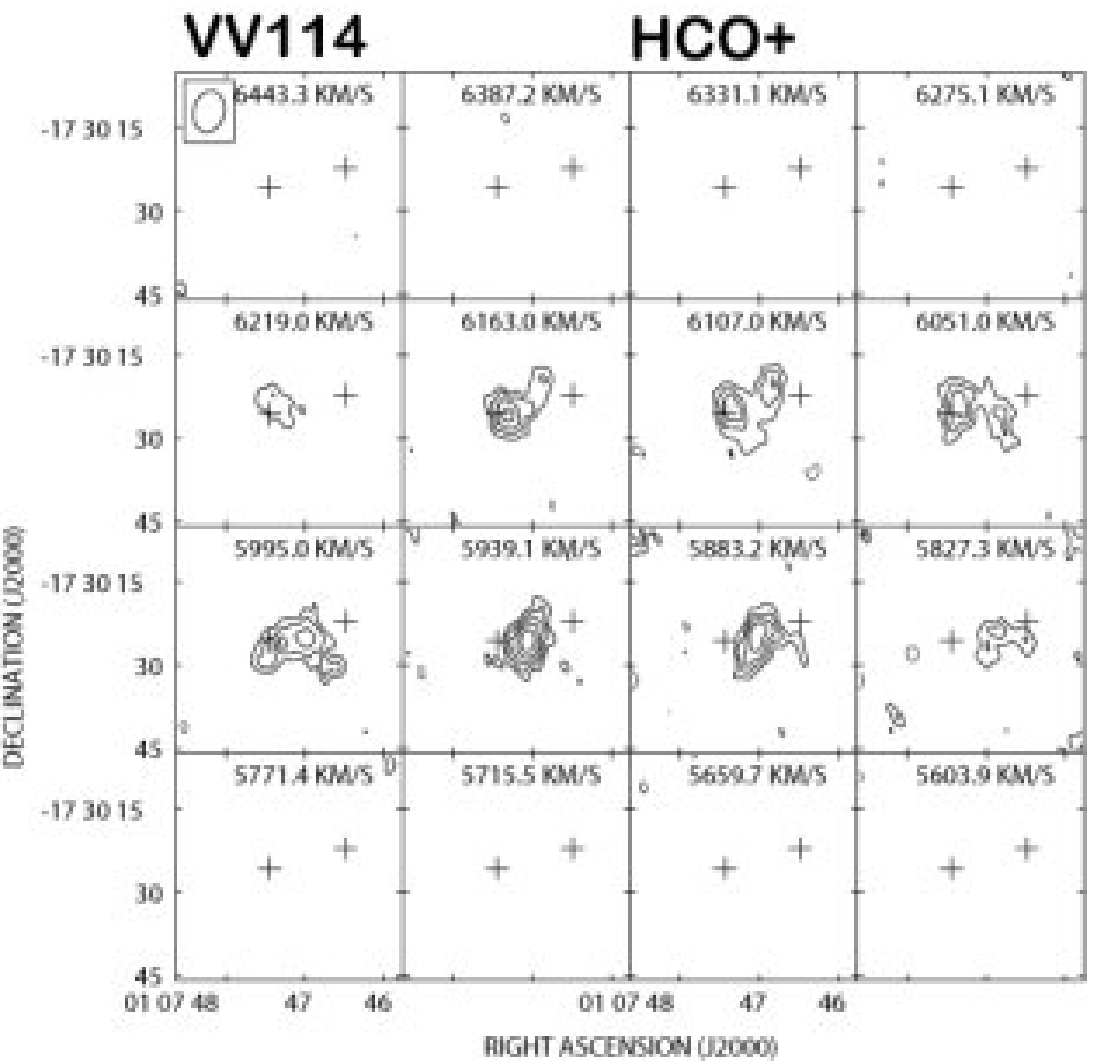}
\caption{
Channel maps of HCN(1--0) and HCO$^{+}$(1--0) emission for VV 114.
{\it Left}: HCN(1--0) emission.
The contours are 3 $\times$ ($-$3, 3, 4, 5) mJy beam$^{-1}$.
The r.m.s. noise level is $\sim$3 mJy beam$^{-1}$.
{\it Right}: HCO$^{+}$(1--0) emission.
The contours are 3 $\times$ ($-$3, 3, 4, 5, 6, 7) mJy beam$^{-1}$.
The r.m.s. noise level is $\sim$3 mJy beam$^{-1}$.
}
\end{figure}

\clearpage

\begin{figure}
\includegraphics[angle=-90,scale=.35]{f5a.eps} 
\includegraphics[angle=-90,scale=.35]{f5b.eps} 
\includegraphics[angle=-90,scale=.35]{f5c.eps} 
\includegraphics[angle=-90,scale=.35]{f5d.eps} 
\end{figure}

\clearpage

\begin{figure}
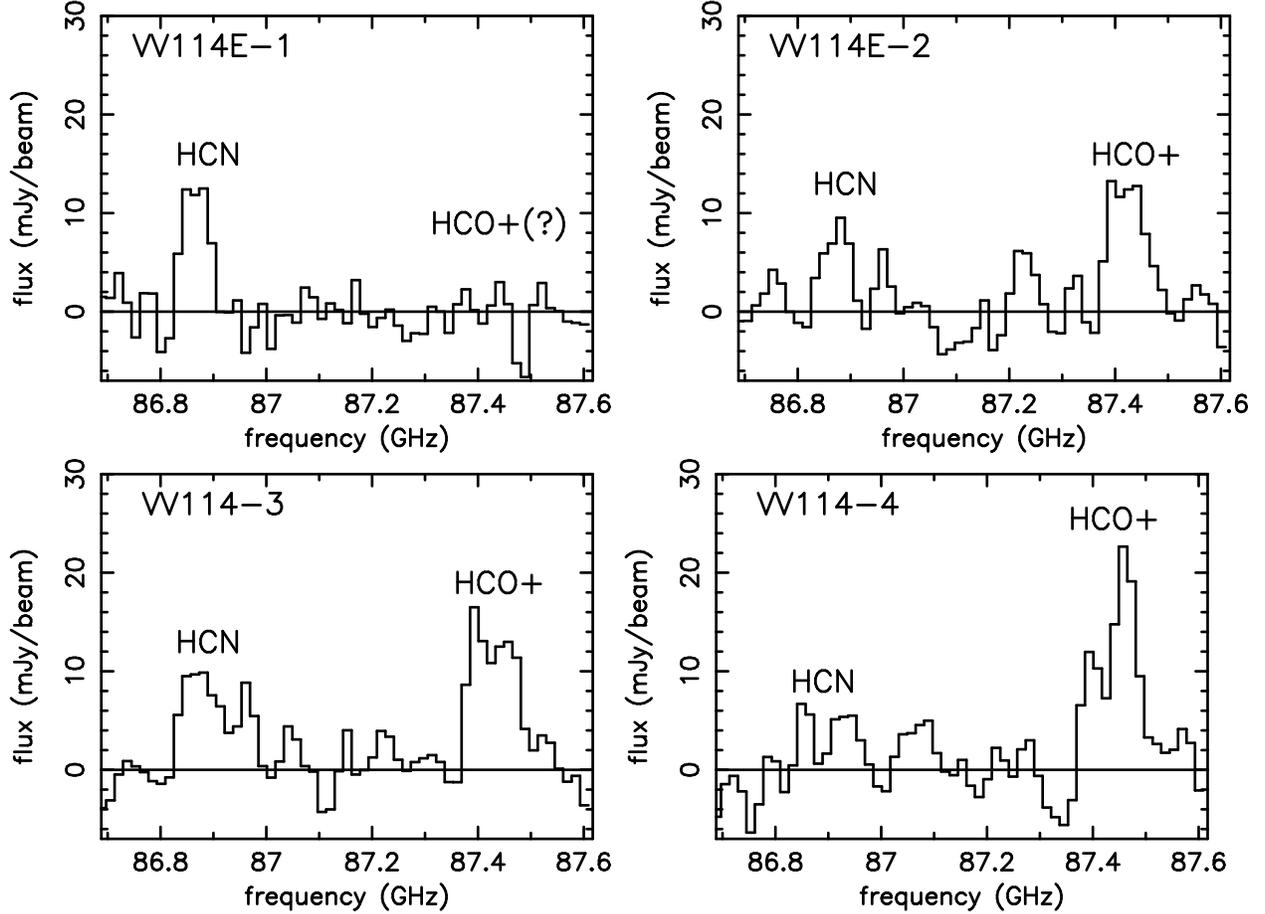

\includegraphics[angle=-90,scale=.35]{f5e.eps} 
\includegraphics[angle=-90,scale=.35]{f5f.eps} 
\includegraphics[angle=-90,scale=.35]{f5g.eps} 
\includegraphics[angle=-90,scale=.35]{f5h.eps} 
\caption{
HCN(1--0) and HCO$^{+}$(1--0) spectra of the observed LIRGs.
The abscissa is the observed frequency in GHz and the ordinate is flux in
mJy beam$^{-1}$. 
For Arp 220 and Mrk 231, the strong continuum emission is subtracted. 
For VV 114, a possible continuum is subtracted.
For IRAS 08572+3915, the continuum is not subtracted, because 
its subtraction introduces a significant uncertainty, given the
faintness of the HCN(1--0) and HCO$^{+}$(1--0) emission lines.
Appropriate spectral binning is made, depending on the signal levels of
the HCN(1--0) and HCO$^{+}$(1--0) emission lines.   
The r.m.s. noise levels per spectral bin in the final spectra 
are $\sim$4 mJy (Arp 220), $\sim$6 mJy (Mrk 231), $\sim$1.5 mJy (IRAS
08572+3915), and $\sim$3 mJy (VV 114).
}
\end{figure}

\begin{figure}
\includegraphics[angle=-90,scale=.35]{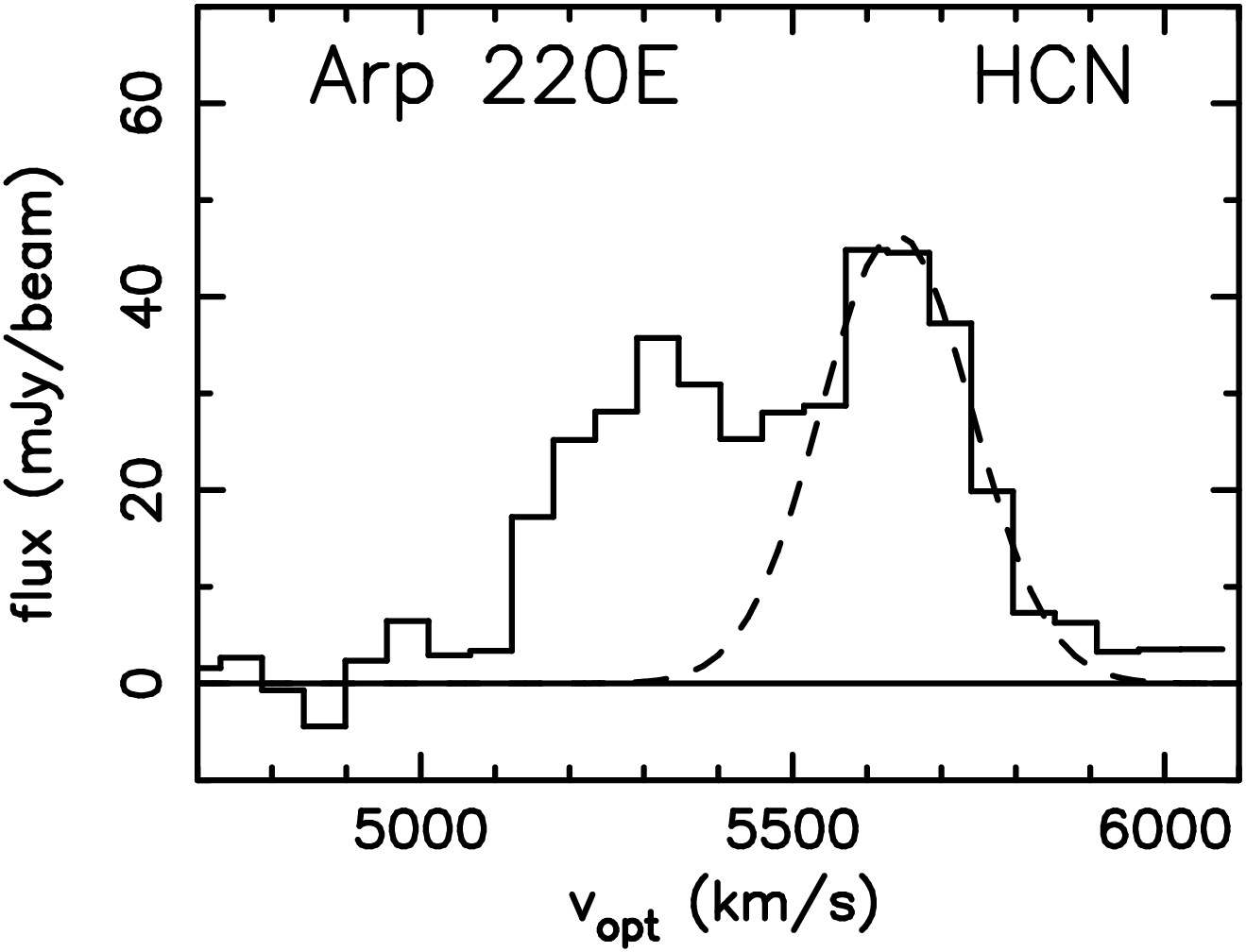}
\includegraphics[angle=-90,scale=.35]{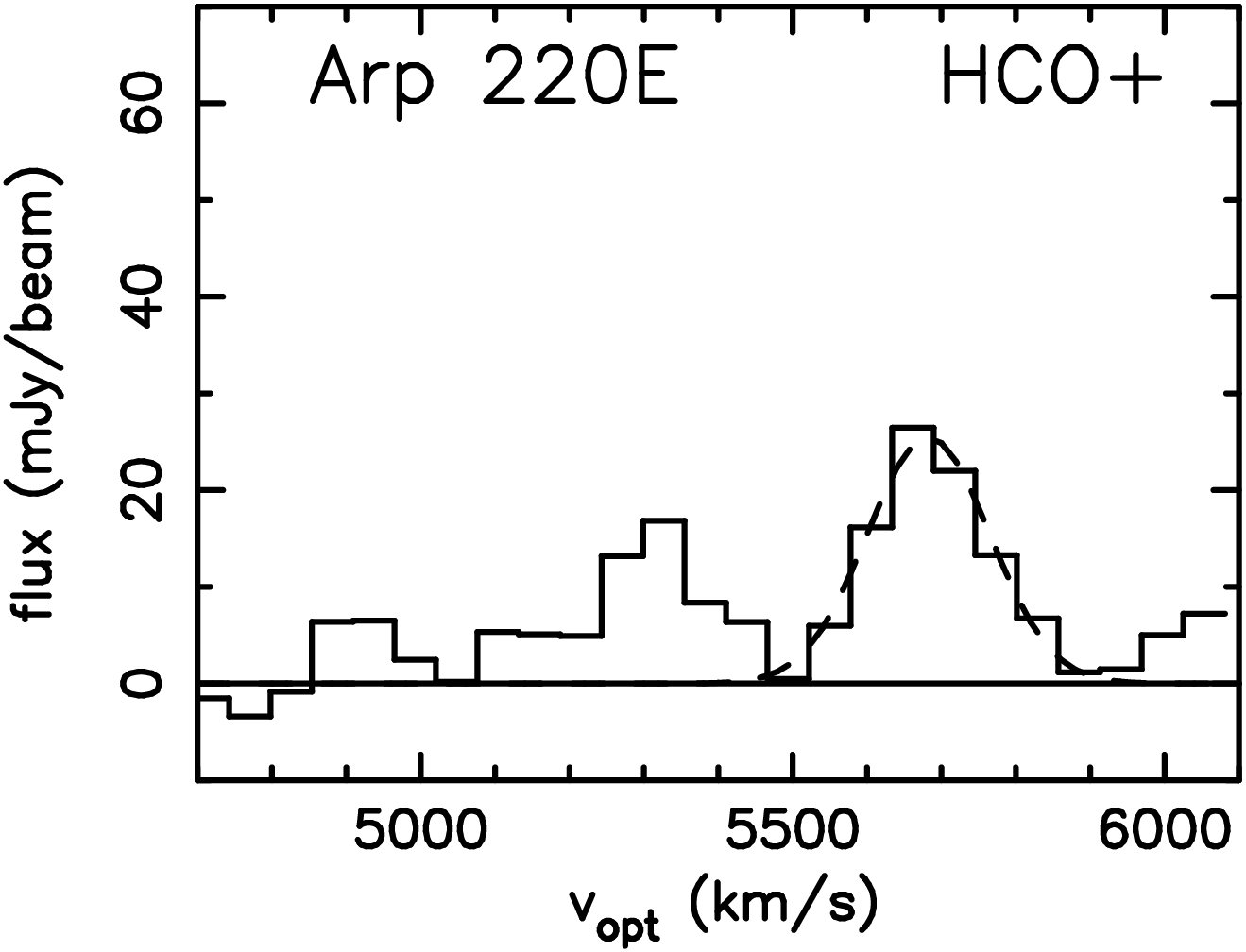}
\includegraphics[angle=-90,scale=.35]{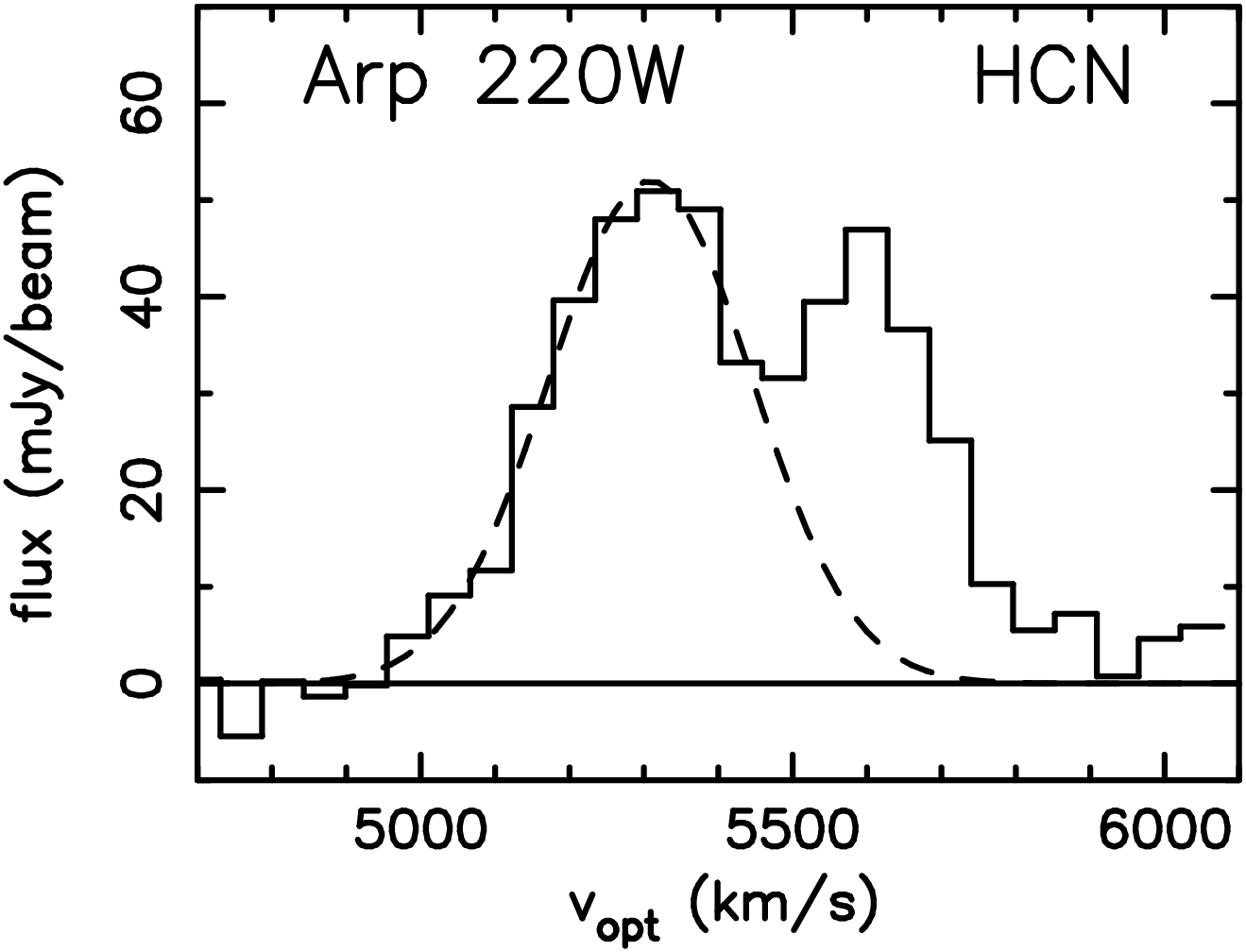}
\includegraphics[angle=-90,scale=.35]{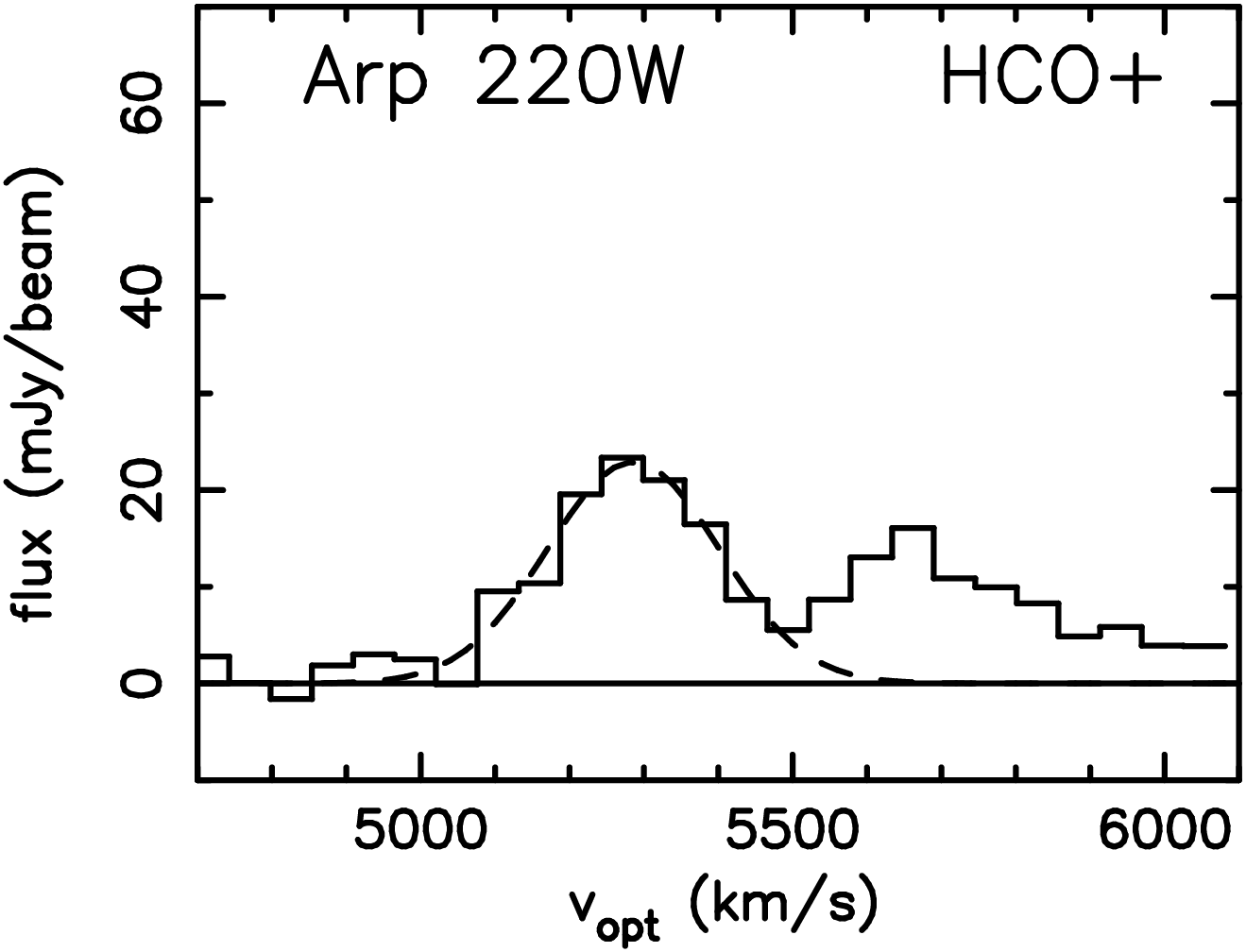}
\includegraphics[angle=-90,scale=.35]{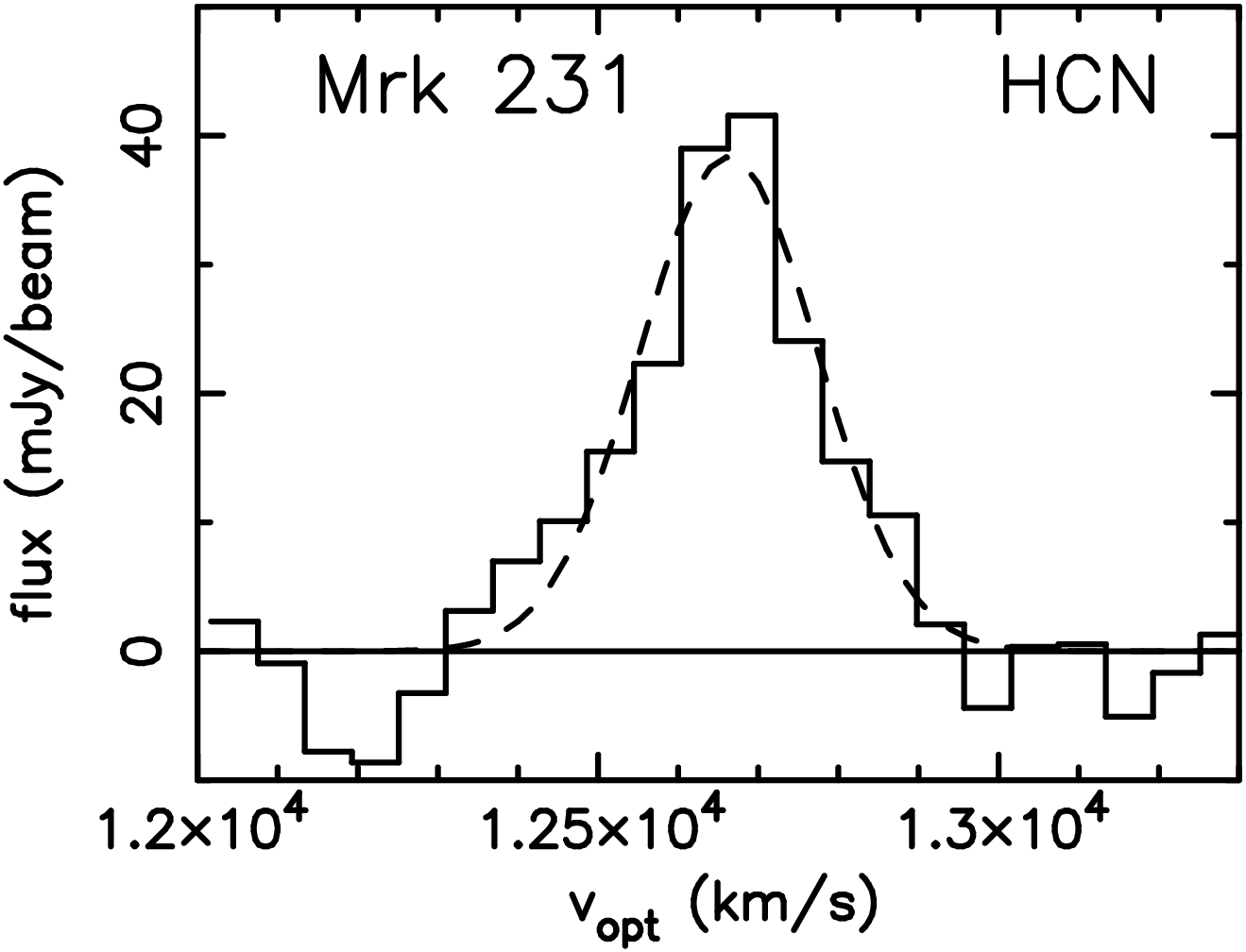}
\hspace{0.7cm}
\includegraphics[angle=-90,scale=.35]{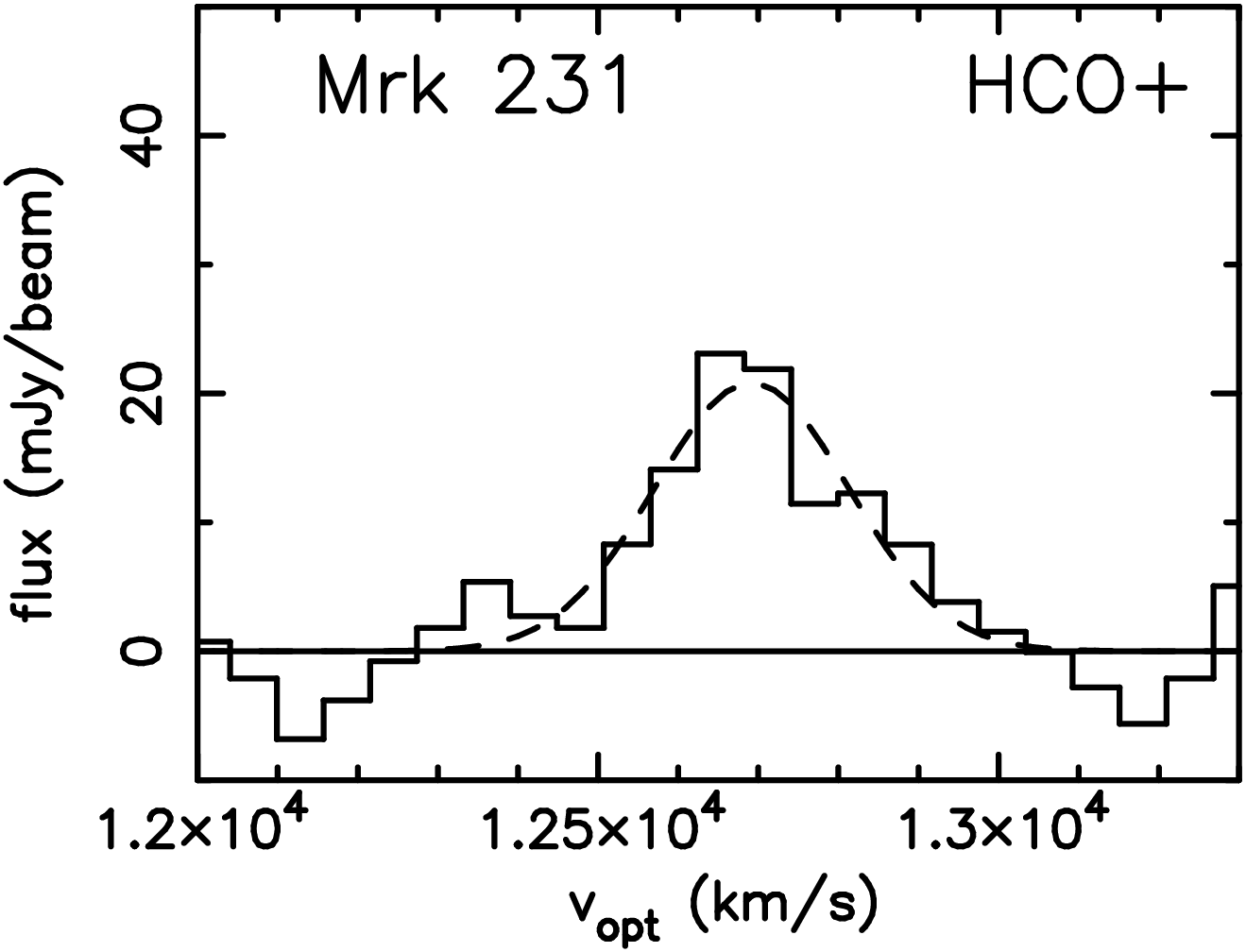}
\end{figure}

\clearpage

\begin{figure}
\includegraphics[angle=-90,scale=.35]{f6g.eps} 
\includegraphics[angle=-90,scale=.35]{f6h.eps}
\includegraphics[angle=-90,scale=.35]{f6i.eps} \\
\includegraphics[angle=-90,scale=.35]{f6j.eps} 
\hspace{0.7cm}
\includegraphics[angle=-90,scale=.35]{f6k.eps} 
\end{figure}

\clearpage

\begin{figure}
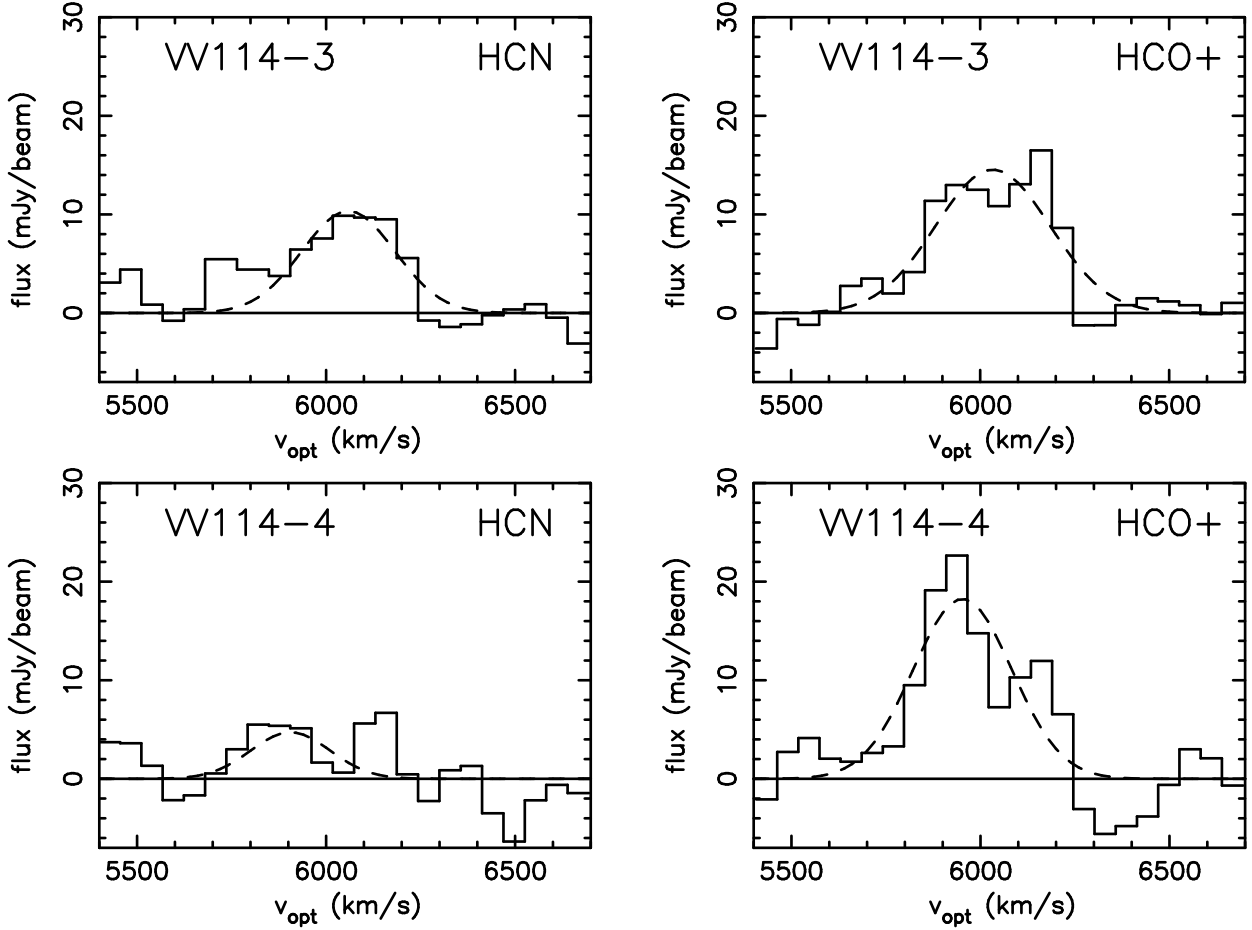

\includegraphics[angle=-90,scale=.35]{f6l.eps} 
\includegraphics[angle=-90,scale=.35]{f6m.eps} 
\includegraphics[angle=-90,scale=.35]{f6n.eps} 
\hspace{0.7cm}
\includegraphics[angle=-90,scale=.35]{f6o.eps} 
\caption{
Gaussian fits to the detected HCN(1--0) and HCO$^{+}$(1--0) emission
lines.  The abscissa is the LSR velocity \{v$_{\rm opt}$ $\equiv$
($\frac{\nu_0}{\nu}$ $-$ 1) $\times$ c\} in km s$^{-1}$ and the
ordinate is flux in mJy beam$^{-1}$.  Single Gaussian fits are
used as defaults and are shown as dashed lines.  
The spectrum of IRAS 08572+3915 is only for the NW nucleus.
For this source, since evidence for double peaks exists, we attempted a
double Gaussian fit, shown as dashed lines. 
A constant continuum is assumed and is set as a free parameter because
the continuum emission was not subtracted from the spectrum 
of IRAS 08572+3915. 
For other sources, the continuum level is set as the zero value. 
The adopted continuum levels are shown as the horizontal solid straight
lines for all sources. 
}
\end{figure}

\clearpage

\begin{figure}
\includegraphics[angle=-90,scale=.35]{f7a.eps} 
\includegraphics[angle=-90,scale=.35]{f7b.eps} 
\includegraphics[angle=-90,scale=.35]{f7c.eps} 
\end{figure}

\begin{figure}
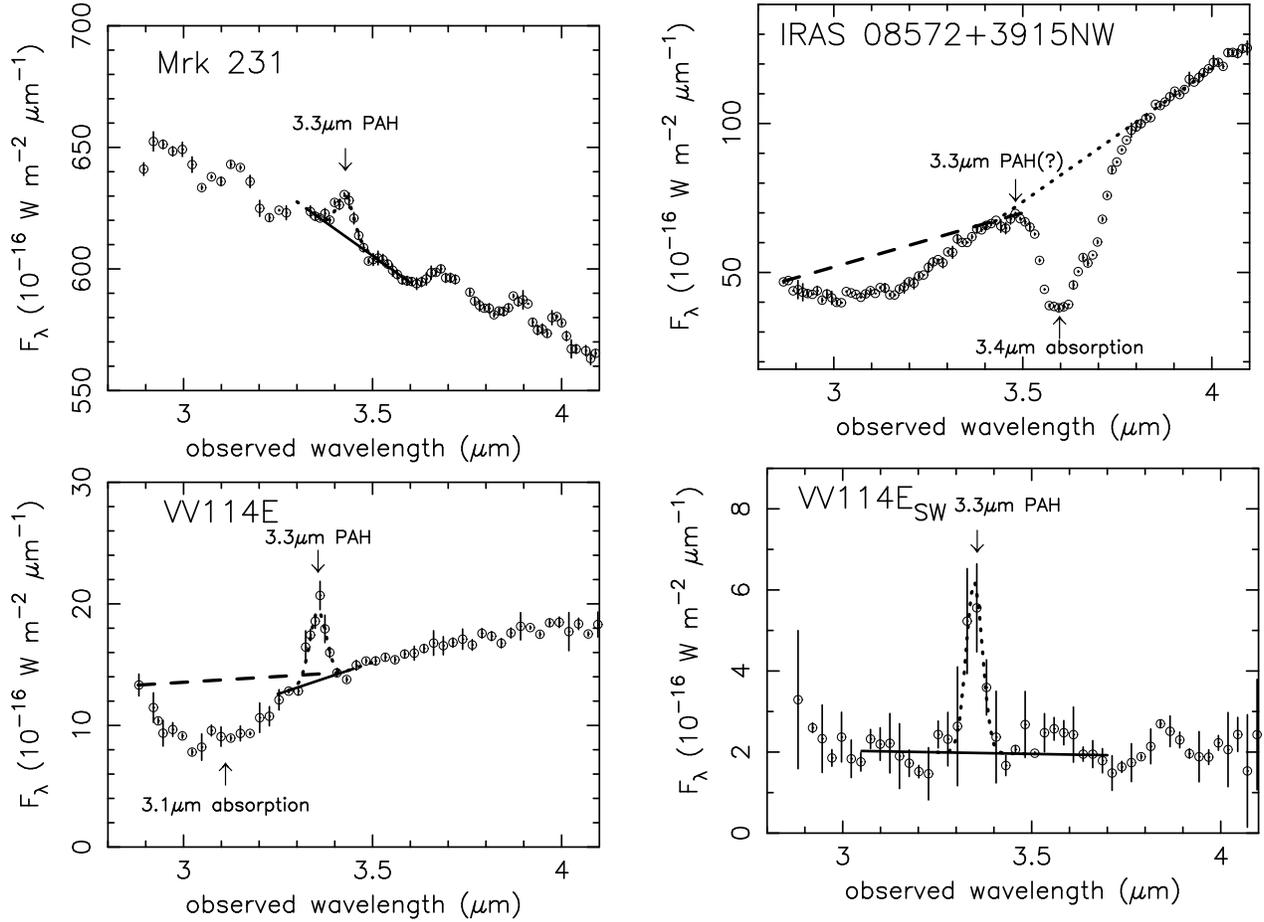

\includegraphics[angle=-90,scale=.35]{f7d.eps} 
\includegraphics[angle=-90,scale=.35]{f7e.eps} 
\includegraphics[angle=-90,scale=.35]{f7f.eps} 
\hspace{0.7cm}
\includegraphics[angle=-90,scale=.35]{f7g.eps} 
\caption{
Newly obtained infrared 2.8--4.1 $\mu$m ($L$-band) spectra of Arp 220, Mrk
231, IRAS 08572+3915NW, and VV 114.
The abscissa is observed wavelength in $\mu$m, and 
the ordinate is F$_{\lambda}$ in 10$^{-16}$ W m$^{-2}$ $\mu$m$^{-1}$. 
Dashed straight line: Adopted continuum level to measure the optical
depth of the 3.1 $\mu$m ice absorption feature.
Dotted line: Gaussian fit to the 3.3 $\mu$m PAH emission feature.
Solid straight line: Adopted continuum level to measure the strength of
the 3.3 $\mu$m PAH emission feature.
Dotted straight line for IRAS 08572+3915 is the adopted continuum level
to measure the optical depth of the 3.4 $\mu$m carbonaceous dust 
absorption feature. 
}
\end{figure}

\begin{figure}
\begin{center}
\includegraphics[angle=-90,scale=.8]{f8.eps}
\end{center}
\caption{\small 
HCN(1--0)/HCO$^{+}$(1--0) (ordinate) and HCN(1--0)/CO(1--0) (abscissa)
ratios in brightness temperature ($\propto$ $\lambda^{2}$ $\times$ flux
density), derived from our NMA/RAINBOW interferometric observations.   
Arp 220, Mrk 231, IRAS 08572+3915, VV 114, and He 2--10 
are plotted as large filled stars with labels.  
Other LIRGs previously observed by \citet{ima04,ink06} and \citet{in06} 
are also plotted as small filled stars.  
Other data points are taken from \citet{koh05}, where sources with
AGN-like (starburst-like) ratios are marked with filled squares (open
circles).
The ratio at the core of the starburst galaxy NGC 253, derived by 
spatially-resolved interferometric data \citep{knu07}, is also plotted 
as a large open triangle. 
For LIRGs and He 2--10, the HCN(1--0)/HCO$^{+}$(1--0)
brightness-temperature ratios are derived directly from our
interferometric data, while the HCN(1--0)/CO(1--0) ratios are derived by
combining our data with CO data in the literature.   
The HCN(1--0)/CO(1--0) ratio in Arp 220 is taken as an upper limit,
because the beam size of CO observations is much smaller.
For all LIRGs' nuclei, the HCN(1--0)/HCO$^{+}$(1--0)
brightness-temperature ratios in the ordinate are those toward the 
cores, where putative buried AGNs are expected to reside in. 
Contamination from extended starburst emission outside the beam sizes 
(Table 3) are totally removed.
}
\end{figure}

\begin{figure}
\begin{center}
\includegraphics[angle=0,scale=1.5]{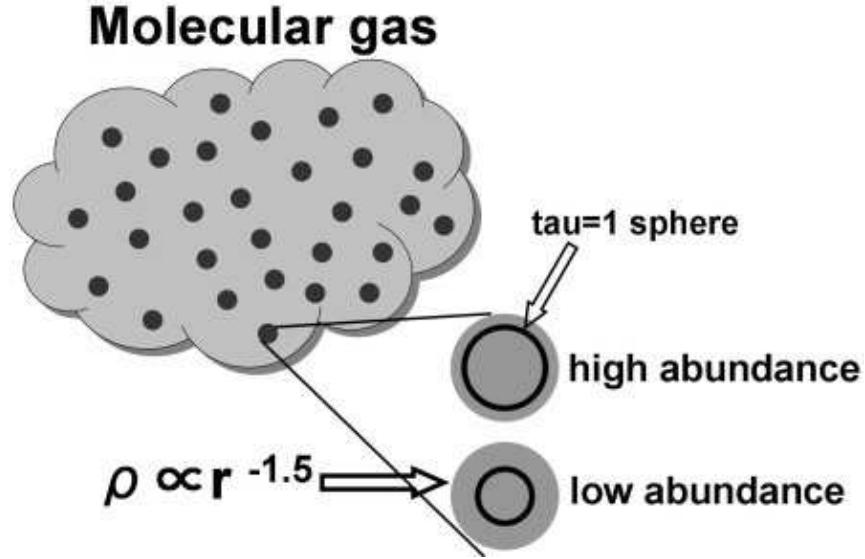}
\end{center}
\caption{
Schematic diagram of molecular gas in the current standard model 
\citep{sol87}.
Molecular gas consists of dense clumps (shown as dark filled circles in
the upper-left panel) with a small volume filling  factor. 
HCN(1--0) and HCO$^{+}$(1--0) lines can be moderately optically
thick for each clump \citep{ngu92}. 
It is very likely that each clump has a decreasing radial density profile
($\propto$ r$^{\alpha}$; $\alpha <$ 0).
If the radial density distribution of r$^{-1.5}$ is assumed
\citep{gie92}, line emission comes mainly from the layer where an
optical depth reaches about unity ($\tau$ $\sim$ 1) \citep{gie92}. 
When a molecular abundance increases (decreases), the $\tau$ $\sim$ 1
layer for certain molecular line moves outward (inward).
Consequently, the surface area of the layer, or its area filling factor 
in molecular gas, becomes larger (smaller), increasing (decreasing) the
flux of that molecular line.  
}
\end{figure}

\begin{figure}
\begin{center}
\includegraphics[angle=-90,scale=.8]{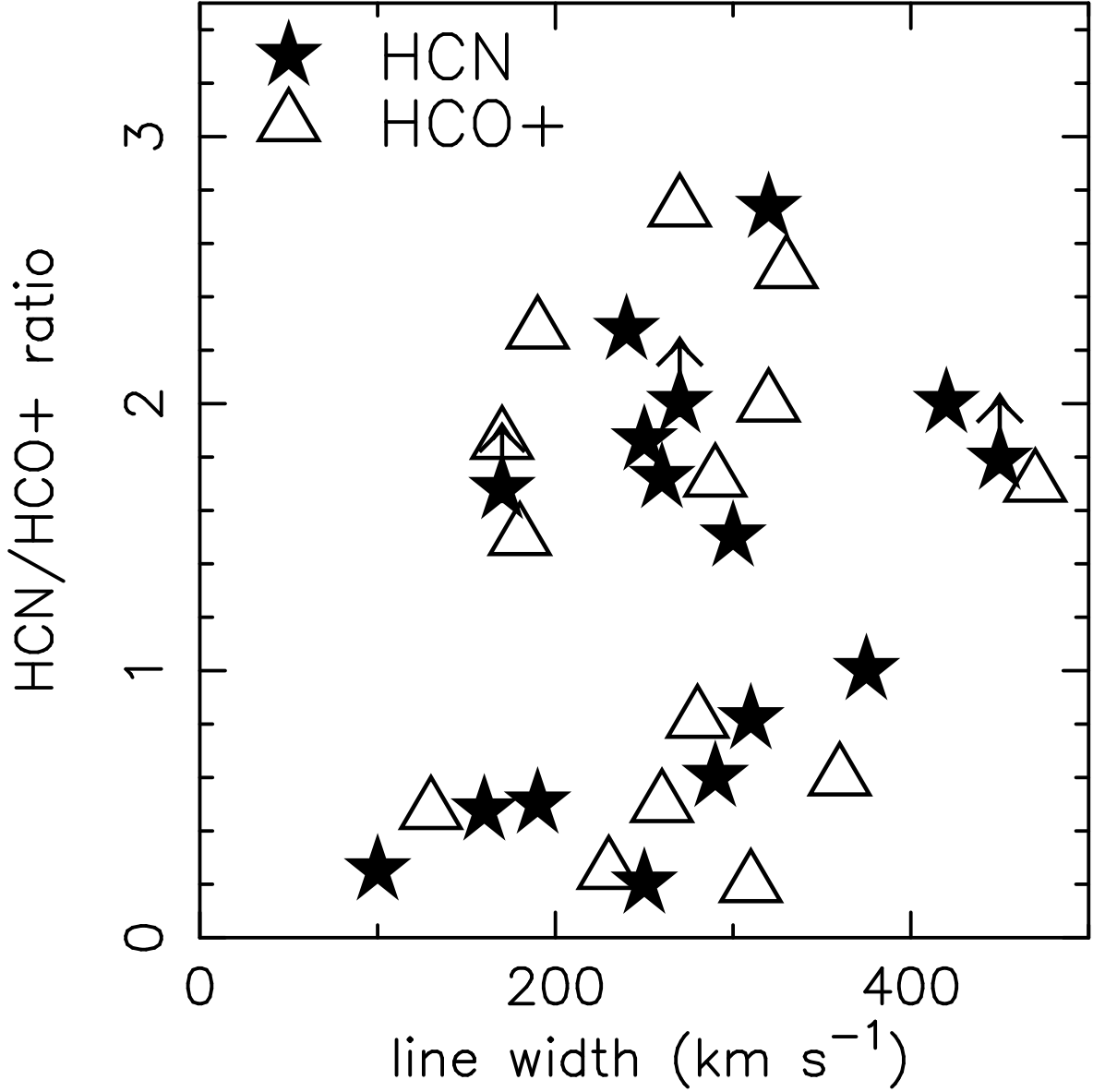}
\end{center}
\caption{
Comparison between FWHM line widths of molecular lines (abscissa) and
HCN(1--0)/HCO$^{+}$(1--0) brightness-temperature ratio (ordinate). 
Filled stars: HCN(1--0) line width.
Open triangles: HCO$^{+}$(1--0) line width.
All LIRGs with available NMA/RAINBOW HCN(1--0) and HCO$^{+}$(1--0) data
are plotted.  
These LIRGs are NGC 4418 \citep{ima04}, UGC 5101, Mrk 273, IRAS
17208$-$0014 \citep{ink06}, Arp 299 A, B, C \citep{in06}, and 
the four LIRGs (This paper).
The HCN(1--0) line width of UGC 5101, and the HCN(1--0) and
HCO$^{+}$(1--0) line widths of IRAS 08572+3915 are for individual
components of the double Gaussian fits (Imanishi et al. 2006b; This
paper).
The HCN(1--0)/HCO$^{+}$(1--0) brightness-temperature ratios 
are also derived for individual components of the double Gaussian fits
(Imanishi et al. 2006b; This paper).   
For Mrk 273 and VV 114E-1, no HCO$^{+}$(1--0) line width information
is available (Imanishi et al. 2006b; This paper). 
The FWHM line widths of NGC 4418 are 250 km s$^{-1}$ [HCN(1--0)] and
170 km s$^{-1}$ [HCO$^{+}$(1--0)], which are estimated from the data of
\citet{ima04}.
}
\end{figure}

\begin{figure}
\begin{center}
\includegraphics[angle=-90,scale=.8]{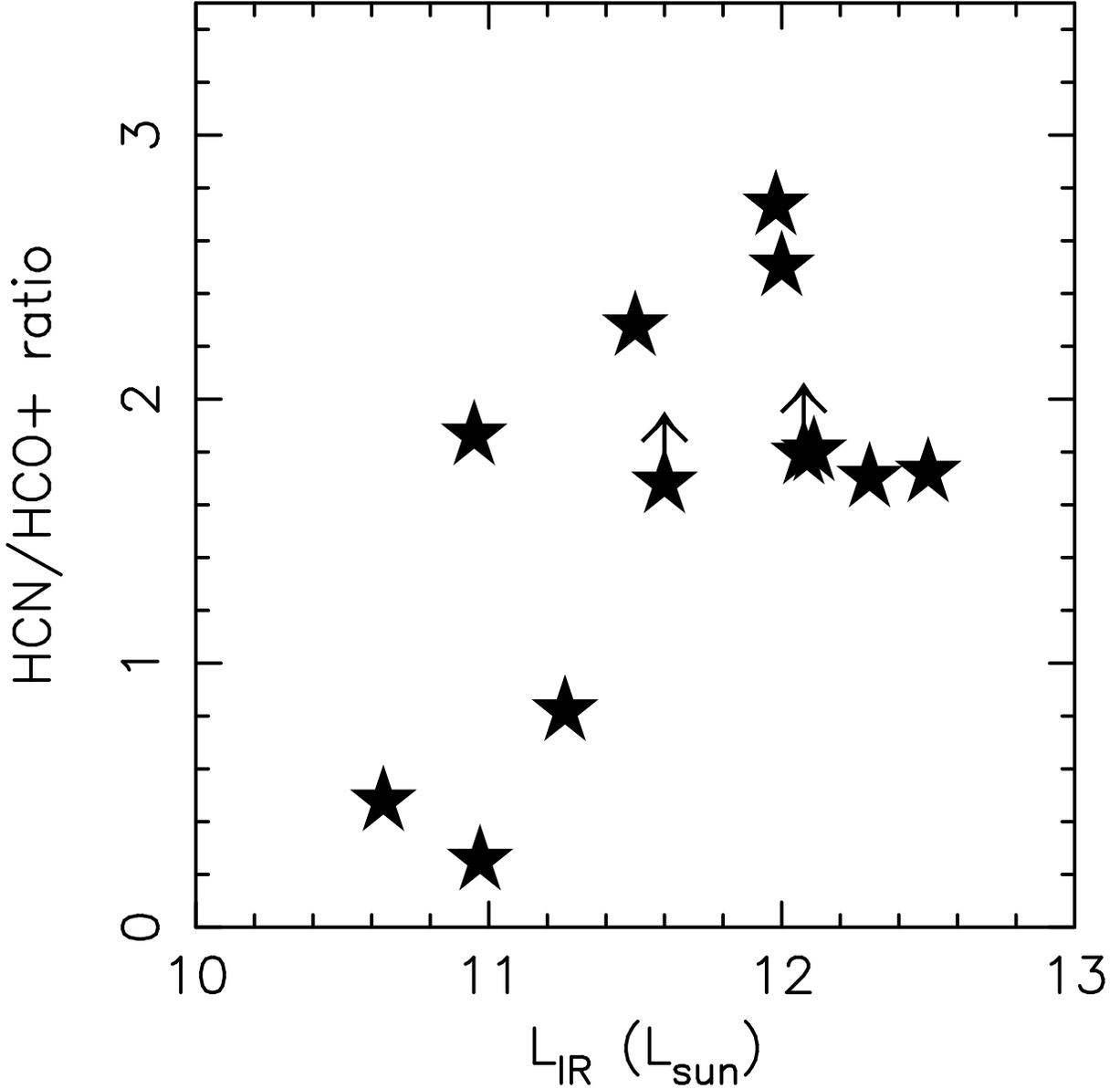}
\end{center}
\caption{
Comparison between {\it IRAS}-measured infrared luminosity in units of
the solar luminosity (abscissa) and HCN(1--0)/HCO$^{+}$(1--0)
brightness-temperature ratio (ordinate).  
The HCN(1--0)/HCO$^{+}$(1--0) ratio in the ordinate is only for 
the core emission within the beam size of interferometric maps, while 
the infrared luminosity in the abscissa traces the emission from the
whole galaxy regions.
For VV 114, we plot the HCN(1--0)/HCO$^{+}$(1--0) brightness-temperature
ratio at VV 114E-1, and assume that the infrared luminosity of VV 114
is dominated by VV 114E.  
For UGC 5101 and IRAS 08572+3915, the HCN(1--0)/HCO$^{+}$(1--0)
brightness-temperature ratios are derived by summing the both components
of the double Gaussian fits. 
For Arp 220, we assume that Arp 220 W and E have, respectively, 75\% and
25\% of the total infrared luminosity \citep{dow07}.
For Arp 299, the infrared luminosities of Arp 299 A, B, C are taken from
\citet{cha02}. 
}
\end{figure}

\end{document}